\documentclass[%
  aps,%
  prd,%
  twocolumn,%
  nofootinbib,%
  floatfix%
  ]%
  {revtex4}
\newcommand*{\myName} {Guido Walter Pettinari}
\newcommand*{\myUni} {University of Portsmouth}
\newcommand*{\myFaculty} {Institute of Cosmology and Gravitation}

\newcommand*{\myAddress} {Institute of Cosmology and Gravitation,
University of Portsmouth,
Dennis Sciama Building,
Burnaby Road,
Portsmouth
PO1 3FX,
United Kingdom
}
\newcommand*{\myTitle} {On the Evidence for Axion-like Particles from Active Galactic Nuclei}

\usepackage{aas_macros}

\usepackage{textcmds}

\usepackage{xspace}

\usepackage[fleqn]{amsmath}
\usepackage{amssymb}

\usepackage[tight,nice]{units}

\usepackage{subfig}

\usepackage{graphicx}

\usepackage{epstopdf}

\usepackage{hyperref}
\hypersetup{%
    colorlinks=true, linktocpage=true, pdfstartpage=3, pdfstartview=FitV,%
    breaklinks=true, pdfpagemode=UseNone, pageanchor=true,%
    pdfpagemode=UseOutlines, plainpages=false, bookmarksnumbered,%
    bookmarksopen=true, bookmarksopenlevel=1, hypertexnames=true,%
    pdfhighlight=/O,
    urlcolor=webbrown, linkcolor=RoyalBlue, citecolor=webgreen, 
    urlcolor=Black, linkcolor=Black, citecolor=Black,
    pdftitle={\myTitle},%
    pdfauthor={\myName, \myUni, \myFaculty},%
    pdfsubject={},%
    pdfkeywords={},%
    pdfcreator={pdfLaTeX},%
    pdfproducer={LaTeX with hyperref}%
}

\usepackage[dvipsnames]{xcolor}

\usepackage{multirow}

\usepackage{caption}
\renewcommand*{\L} {\mathcal{L}}  
\newcommand*{\sci} [2] {\ensuremath{{#1} \times 10^{#2}}} 
\newcommand*{\ie} {\emph{i.\,e.}\xspace}
\newcommand*{\eg} {\emph{e.\,g.}\xspace}
\newcommand*{\PDF} {\text{PDF}\xspace}
\newcommand*{\PDFs} {\text{PDFs}\xspace}
\newcommand*{\CDF} {\text{CDF}\xspace}
\newcommand*{\CDFs} {\text{CDFs}\xspace}
\newcommand*{\atan} {\tan^{-1}}
\newcommand*{\logten} {\log_{10}}
\newcommand*{\abs} [1] {|{#1}|}

\newcommand*{\Xray} {X-ray\xspace}
\newcommand*{\gray} {$ \gamma $-ray\xspace}
\newcommand*{\Xrays} {X-rays\xspace}
\newcommand*{\grays} {$ \gamma$-rays\xspace}
\newcommand\ion[2]{#1$\;${\scshape{#2}}}  
\newcommand*{\ROSAT} {\emph{ROSAT}\xspace}

\newcommand*{\avg} [1] {\ensuremath{\langle #1 \rangle}}

\renewcommand{\vec}[1]{\boldsymbol{#1}} 

\newcommand*{\mathlineskip}{\\[0.3cm]}
\newcommand*{\imagewidth} {0.98\linewidth}
\newcommand*{\myfloatalign}{\centering}
\DeclareMathSymbol{\GammaIt}{\mathalpha}{letters}{"00}

\newcommand*{\tref} [1] {Table \ref{#1}\xspace}

\newcommand*{\sref} [1] {Sec.\ \ref{#1}\xspace}

\newcommand*{\eref} [1] {Eq.\ \eqref{#1}\xspace}

\newcommand*{\fref} [1] {Figure \ref{#1}\xspace}

\newcommand*{\cref} [1] {Chapter~\ref{#1}\xspace}

\defcitealias{burrage:2009a}{BDS}
\newcommand*{\BDS}{\citetalias{burrage:2009a}\xspace}
\newcommand*{\meff}{\ensuremath{m_{\text{eff}}}\xspace}
\newcommand*{\mphi}{\ensuremath{m_{\phi}}\xspace}
\newcommand*{\ompl}{\ensuremath{\omega_{\text{pl}}}\xspace}
\newcommand*{\Lhi}{\ensuremath{Y}\xspace}
\newcommand*{\Llow}{\ensuremath{X}\xspace}
\newcommand*{\Ecrit}{\ensuremath{E_{\,\text{crit}}}\xspace}
\newcommand*{\Gaussmodel} {Gaussian model\xspace}
\newcommand*{\ALPmodel} {ALP-mixing model\xspace}
\newcommand*{\Shotmodel} {Shot-noise model\xspace}
\newcommand*{\ALP}{\text{ALP}}

\newcommand*{\ang} [1] {\unit[#1]{\text{\AA}}}
\newcommand*{\kev} [1] {\unit[#1]{keV}}

\newcommand*{\Lang} [1] {\ensuremath{L_{\,\ang{#1}}}\xspace}
\newcommand*{\Lkev} [1] {\ensuremath{L_{\,\unit[#1]{keV}}}\xspace}
\newcommand*{\sdssxmm} {SDSS/XMM-Newton\xspace}
\newcommand*{\oscprob}{\ensuremath{P_{\gamma\leftrightarrow\phi}}\xspace}
\newcommand*{\mixing}{photon-ALP mixing\xspace}
\newcommand*{\coupling}{photon-ALP coupling\xspace}
\newcommand*{\NoverN}{\ensuremath{\nicefrac{N}{\avg{N}}}\xspace}
\newcommand*{\var} [1] {\ensuremath{\sigma_\text{#1}^2}}
\newcommand*{\KS} {Kolmogorov-Smirnov\xspace}
\newcommand*{\rtest} {$ r $-test\xspace}
\newcommand*{\rstat} {$ r $-statistic\xspace}
\newcommand*{\rdist} {$ r $-distribution\xspace}
\newcommand*{\rdists} {$ r $-distributions\xspace}

\newcommand*{\nagn} {\ensuremath{N_{\text{AGN}}}\xspace}
\newcommand*{\Pmix} {\ensuremath{P_\text{mix}}\xspace}

\newcommand*{\sigmain} {\ensuremath{\sigma_{\text{in}}}\xspace}
\newcommand*{\sigmamix} {\ensuremath{\sigma_{\text{mix}}}\xspace}
\newcommand*{\sigmatot} {\ensuremath{\sigma_{\text{tot}}}\xspace}

\newcommand*{\sigmakev} [1] {\ensuremath{\sigma_{\kev{#1}}}}

\newcommand*{\nagnBDSa} {77\xspace}
\newcommand*{\nagnBDSb} {203\xspace}
\newcommand*{\nagnFull} {340\xspace}
\newcommand*{\nagnHigh} {320\xspace}
\newcommand*{\nagnLowG} {20\xspace} 

\newcommand*{\nresamp}{$ 10^5 $\xspace}

\newboolean{catalogswithnumbers}
\setboolean{catalogswithnumbers}{false} 
\ifthenelse{\boolean{catalogswithnumbers}}{
   \newcommand*{\BDSa} {BDS-\nagnBDSa}
   \newcommand*{\BDSb} {BDS-\nagnBDSb}
   \newcommand*{\Full} {Full-\nagnFull}
   \newcommand*{\HighG} {High$\Gamma$-\nagnHigh}
}{ 
   \newcommand*{\BDSa} {BDS-\nagnBDSa}
   \newcommand*{\BDSb} {BDS-\nagnBDSb}
   \newcommand*{\Full} {Full\xspace}
   \newcommand*{\HighG} {High-$\Gamma$\xspace}
}
\newcommand*{\nsimsrtest} {\ensuremath{10,\!000}\xspace}
\newcommand*{\nsimsgoftest} {\ensuremath{50,\!000}\xspace}
\newcommand*{\pmixhigh} {\ensuremath{0.7}\xspace}
\newcommand*{\pmixmiddle} {\ensuremath{0.4}\xspace}
\newcommand*{\pmixlow} {\ensuremath{0.1}\xspace}

\begin{document}

\title{\myTitle}
\author{\myName}
\email{Guido.Pettinari@port.ac.uk}
\author{Robert Crittenden}
\address{\myAddress}
\begin{abstract}
\citet*{burrage:2009a} recently suggested exploiting the correlations between high and low energy luminosities of astrophysical objects to probe possible mixing between photons and axion-like particles (ALP) in magnetic field regions.  They also presented evidence for the existence of ALP's by analyzing the optical/UV and \Xray monochromatic luminosities of AGNs.   We extend their work by using the monochromatic luminosities of $ \nagnHigh $ unobscured Active Galactic Nuclei from the Sloan Digital Sky Survey/Xmm-Newton Quasar Survey \citep{young:2009b}, which allows the exploration of  $ 18 $ different combinations of optical/UV and \Xray monochromatic luminosities.   However, we do not find compelling evidence for the existence of ALPs.   Moreover, it appears that the signal reported by \citeauthor{burrage:2009a} is more likely due to \Xray absorption rather than to photon-ALP oscillation.
\end{abstract}

\maketitle


\section{Introduction}
\label{sec:introduction}
Very light scalar fields could have had a significant impact on cosmology, potentially acting as dark matter (\eg, axions or axion-like particles) or explaining the recent accelerated expansion (\eg, a quintessence field).   In both cases, their roles are primarily gravitational, either to provide additional gravitational clustering on galactic or cluster scales, or to drive the overall expansion of the Universe.   However, this does not rule out non-gravitational interactions, which should also exist if not explicitly forbidden by some symmetry \citep{carroll:1998a}.

The strengths of such interactions are well constrained for axion-like particles (ALPs). These interactions could lead to more efficient stellar cooling, and the limits from solar axions observed on Earth constrain the couplings to be $ g \lesssim \unit[10^{-10}]{GeV^{-1}} $ \citep{amsler:2008a}. Additionally, ALPs can be emitted from the core of supernovae at a significant rate \citep{raffelt:2008a}; the lack of evidence for such outburst from SN 1987A yields $ g \lesssim \unit[10^{-11}]{GeV^{-1}} $ for a very light ALP ($ m \lesssim \unit[10^{-9}]{eV} $) \citep{amsler:2008a}.  Similarly in the quintessence case, such interactions between photons with a slowly rolling field would lead to time variations in the fine structure constant which could be observed in stellar lines \citep{carroll:1998a}.

One way of avoiding these constraints is the so-called chameleon model \citep{khoury:2004a}, where non-minimal couplings to gravity lead to the effective mass or coupling of the scalar field being dependent on the local mass density.  In this way, many of the constraints on the mass of the scalar fields can be satisfied, while still allowing for a significant interaction strength in regions where the Universe is less dense \citep{brax:2007a, mota:2007a}.

It is worth trying to constrain couplings of the axion-like particles in low density regions where they are not masked by chameleon effects.  Typical interactions couple the axion-like particle to two photons; this can lead to photons decaying into axions as they pass through magnetic fields.   If the probability is small, this leads to `tired light' scenarios, where objects at large distances are progressively dimmer than expected.  These ideas have been investigated in the context of Type Ia supernovae, where the effect is similar to that of grey dust \citep{csaki:2002a, csaki:2002b, bassett:2004b, bassett:2004a} but still cannot explain their apparent dimming without introducing cosmic acceleration (see Ref.\ \citep{mirizzi:2008a} and references therein).

In the opposite strong-mixing regime, the photons could convert to axions rarely, but with high probability, as they pass through the magnetic fields around a galaxy or cluster.  On average, the mixing will result in one third of the photons being converted into ALP's, but the exact amount will depend on the magnetic field orientations along the individual photon paths.  Given the relatively short coherence lengths associated with such magnetic fields, the mixing is expected to vary from source to source.

The average suppression of such strong mixing is difficult to detect without well calibrated sources at cosmological distances. However \citet*{burrage:2009a} (hereafter, \citetalias{burrage:2009a}) recently proposed using the distribution of the fluxes of cosmological sources as a means of constraining the mixing.   Since the mixing is dependent of the photon energy, the low energy fluxes are used to calibrate the brightness of the sources.   After considering a number of potential sources, \citetalias{burrage:2009a} analysed the distribution of X-ray fluxes of AGN, normalised by their optical fluxes and found significant evidence of such mixing, up to the 5$\sigma$ level in their most recent analysis \cite{burrage:2009c}.

Such strong evidence for ALP mixing is tantalizing, and the purpose of this paper is to re-examine and extend their analysis with larger data sets to see how robust the signal is.   In section II, we review the dynamics of the mixing model and the expected signal.  In section III, we examine a number of possible ways to evaluate the statistical significance of such a signal, and in section IV we apply these methods, re-examining the original claim of \citetalias{burrage:2009a} and then extending the analysis to the larger data set of \citet{young:2009b}, before concluding in section V.

\section{Model Assumptions}
\label{sec:model_assumptions}
\subsection{Scatter from mixing to Axion-like particles}
For our purposes, an axion-like particle (ALP) is a light but massive scalar or pseudo-scalar field that couples with the kinetic term of the Lagrangian of the photons. Depending on whether the ALP is a scalar or a pseudoscalar, this interaction term has one of the following forms:
\begin{equation*}
  \begin{aligned}
	&  \L^{^S}_{\text{int}} = \frac{\phi}{M} \: (\vec{B}^2 - \vec{E}^2)
  \mathlineskip
	&  \L^{^{PS}}_{\text{int}} = \frac{\phi}{M} \: \vec{E}\cdot\vec{B} \;,
  \end{aligned}	
\end{equation*}
where $ \phi $ is the axion field and $ M \equiv 1/g $ sets the scale of the strength of the \coupling. As a result, when a photon of energy $ E $ travels through a magnetic domain of length $ L $ and intensity $ B $, there is a non-zero probability that it oscillates into an ALP \citep{hooper:2007a, sikivie:1983a, raffelt:1988a}:
\begin{equation}
\label{eq:alp_oscillation_prob}
    \oscprob =
      \frac{1}{1 + \Ecrit^{\,2} / E^{\,2}}
      \sin^2
      \left(
        \frac{\mu \, L}{2} \, \sqrt{1 + \frac{\Ecrit^{\,2}}{E^{\,2}}}
      \right)
\end{equation}
where $ \Ecrit = \mphi^2 / 2\mu $ is the characteristic energy scale, $ \mu \equiv B/M $, and $ \mphi $ is the ALP mass.

We are most interested in ALP-Photon mixing in an astrophysical context where the propagating medium is an electron plasma. Therefore, we have to substitute an effective mass for the mass of the ALP, $\mphi \rightarrow \meff$, where,
\begin{equation*}
   \meff^2 \equiv |\mphi^2 - \ompl^2 - \epsilon\,\mu^2| \;;
\end{equation*}
here, $ \ompl^2 = 4\pi\alpha_{\text{em}}n_e/m_e $ is the plasma frequency, $ \alpha_{\text{em}} $ is the fine structure constant, $ n_e $ is the free electron number density and $ m_e $ is the mass of the electron. The parameter $ \epsilon $ can be either $ 1 $ (scalar ALP) or $ 0 $ (pseudo-scalar ALP); in the following discussion we shall always assume $ \epsilon = 0 $, \ie a pseudo-scalar axion-like particle.  In the typical environments we are interested in, the free electron densities are of order $ \unit[10^{-2} - 10^{-3}]{cm^{-3}} $, leading to plasma frequencies of order $ \unit[10^{-11}]{eV} $.   For very light masses ($ \mphi < \unit[10^{-12}]{eV} $), the effective mass will be dominated by this plasma frequency.

Here, we focus on mixing which would occur as photons transverse a typical intra-cluster medium, where there are magnetic domains of coherent length $ L \sim \unit[1-100]{Kpc} $ and intensity $ B \sim \unit[1-10]{\mu G} $ \citep{carilli:2002a}.   Passing through a whole cluster of length \unit[1]{Mpc}, the photons will encounter a number ($ N \gg 1 $) of independent magnetic domains.   Initially we shall assume that every light path either crosses a cluster while travelling towards us or is originated inside a cluster. This assumption is optimistic and we shall discuss what happens when it is relaxed in \sref{sec:stat_analysis}.

We are interested in the \emph{strong mixing limit} and when the mixing is independent of the photon energy.   Strong mixing occurs in a single domain when $BL/2M \ge 1$; that is, if the magnetic fields are sufficiently strong, are coherent over a large enough region or the coupling to the ALP is high enough.   For typical cluster magnetic fields, strong mixing in a single coherent region requires $M < \unit[10^{11}]{GeV}$.   However, even if the mixing probability is small in a given region, strong mixing over the whole path still occurs as long as $ N\,\oscprob = N \, (BL/2M)^2 \gg 1 $.   If $BL/2M \ge 1$ , energy independence holds if $E \gg \Ecrit$;  for weaker mixing, the energy independence extends to lower energies ($E \gg \Ecrit (BL/2M)$).

\citetalias{burrage:2009a} estimate that the frequency independent and strong mixing limits are both reached when $ E \gtrsim \unit[0.3-3]{keV} $ if one assumes $ \mphi \lesssim \unit[10^{-12}]{eV} $, $ M \lesssim \unit[\sci{3}{11}]{GeV} $ and typical properties for the intra-cluster medium.   Thus, one expects the effects of mixing to be most significant for \Xrays and \grays, and be relatively small for softer photons such as those in the optical or UV bands.   The test we describe will exploit this, by taking the optical luminosities as a direct indicator of the true luminosity in order to normalise the luminosities in harder bands, where mixing could be significant.

\begin{figure}[htbp]
    \myfloatalign
    \includegraphics[width=\imagewidth]{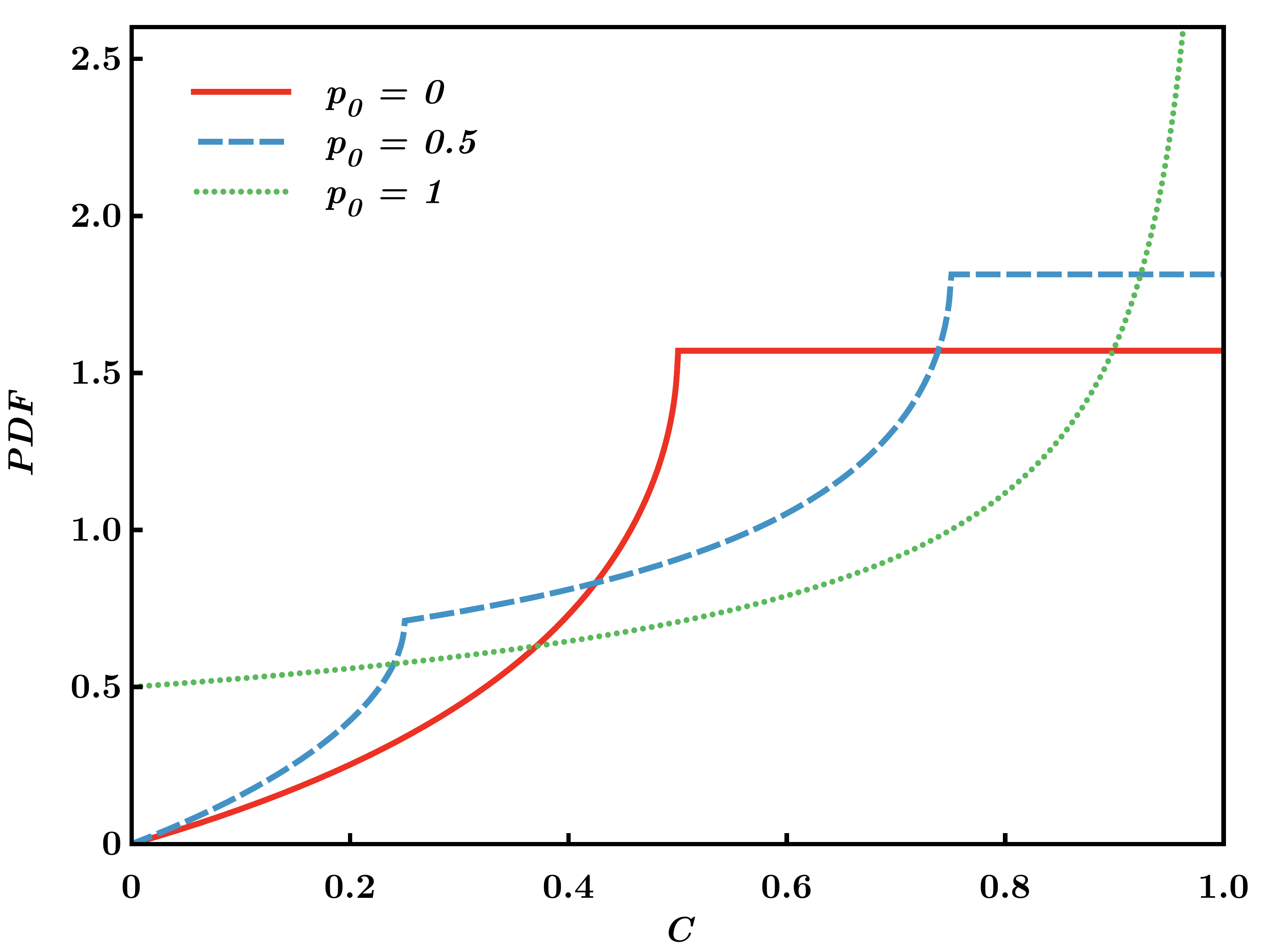}
    \caption{
        Probability distribution function of $ C $, the ratio between the photon and the total intensity, when $ p_0 = 0., \, 0.5, \, 1. $.
        }
    \label{fig:pdf_c}
\end{figure}

In the strong mixing and frequency independent regimes, \citetalias{burrage:2009a} found that, beginning with a pure photon beam, its intensity along a given line of sight will be decreased by a random factor given by:
\begin{equation}
\label{eq:def_c}
C \equiv {I_{\gamma} \over I_{tot}} = 1 - \frac{1+p_0}{2} K_1^2 - \frac{1-p_0}{2}K_2^2,
\end{equation}
where $p_0$ is the initial degree of polarisation and  $K_1$ and $K_2$ are uniformly distributed random variables over the interval $[-1,1].$
The resulting probability distribution of this ratio is given by
\begin{equation}
\label{eq:pdf_c}
	\begin{aligned}
	    f_C (c; p_0) =
	        & \frac{1}{\sqrt{1-p_0^2}}
	        \left\{
	            \atan
	            \left[
	                \sqrt{a}
	                \left(
	                    1 - \frac{2c_+}{1+p_0}
	                \right) ^ {-1/2}
	            \right]
	        \right. \mathlineskip
	            &
	        \left.
	            - \atan
	            \left[
	                \sqrt{a}
	                \left(
	                    1 - \frac{2c_-}{1-p_0}
	                \right)^{1/2}
	            \right]
	        \right\}
	\end{aligned}
\end{equation}
where $ a \equiv ( 1-p_0 ) / ( 1+p_0 ) $, $ c_\pm \equiv \text{min} ( c, (1\pm p_0)/2 ) $ and $ p_0 \in [0,1) $ is the amount of linear polarisation%
\footnote{For details on the polarisation induced by \mixing, refer to \citet{burrage:2009b}.}.  This unusual \PDF is shown in \fref{fig:pdf_c} for three different values of $ p_0 $ . The expectation value of $ C $ given by \eref{eq:pdf_c} is independent of $ p_0 $ and amounts to $ \bar{C} = 2/3 $, meaning that, on average, one-third of emitted photons that cross $ N \gg 1 $ magnetic domains is turned into axions. Its standard deviation increases with $ p_0 $ and is in the range 0.2 -- 0.3.

The curves are perhaps simplest to understand in the fully polarised case, where $K_2$ drops out of Eq. \ref{eq:def_c}, leaving the \PDF described by the Jacobian of the transformation from $K_1$ to $C$, giving $f_C(c)  = \frac{1}{2} (1-c)^{-1/2}$.    In the other cases, one must marginalise over the residual degree of freedom; the hard boundaries on the distributions of the $K$'s lead to the features at $c = (1\pm p_0)/2$, as seen in \fref{fig:pdf_c}.

\subsection{Other sources of scatter}
\label{sec:scatter_models}
If we had perfectly calibrated sources, where we knew their distances, it would be straightforward to tell whether they were dimmed due to \mixing with some scalar field.  In the absence of such calibrated sources, \citetalias{burrage:2009a} proposed to exploit empirical relations between the luminosities in different bands of certain classes of astrophysical objects.   The basic idea is that we have observations of objects in a large range of frequency bands; low frequency luminosities are assumed not to be affected by mixing, and so are taken to be an indicator of the true high frequency luminosity of the object, assuming some empirical relation.   We then examine how the observed high frequency luminosities relate to that predicted; we focus on the characteristic scatter caused by the \mixing, as any average decrease in the luminosity is absorbed in the empirical relations.

We require a relation between a quantity that is affected by \mixing ($\Lhi$), such as \Xray or \gray energy, and one that is unaffected ($\Llow$), such as the optical luminosity or some other low energy feature of the light curve.   A number of such empirical laws exist for AGNs \citep{zamorani:1981a, avni:1982a, vignali:2003a, strateva:2005a, steffen:2006a, just:2007a, gibson:2008a, young:2009b}, Blazars \citep{bloom:2007a, xie:1997a} and \gray bursts \citep{schaefer:2007a}. \citetalias{burrage:2009a} focused on those in the form of a power law:
\begin{equation}
\label{eq:plain_empirical_law}
    \logten(\Lhi) = a + b \logten(\Llow) \;;
\end{equation}
here, the definition of $ \,\Lhi $ and $ \Llow $ depends on the empirical law we are considering. If \mixing occurs, however, we never observe $ \Lhi $ but rather its \qq{dimmed} counterpart $ C \, \Lhi $ (the same does not apply for $ \Llow $ which is assumed to be unchanged by \mixing.)

Even in the absence of \mixing, it is unreasonable to assume that low frequency and high frequency luminosities are perfectly correlated, as there could be many factors affecting these luminosities which vary from object to object.   The origin of this intrinsic scatter depends on the physics of the emission of the different energy photons, which varies according to the type of object under consideration.  (See \sref{sub:problem_xray_absorption} for some specific examples in the AGN case.)   The intrinsic scatter is usually assumed to be a Gaussian distributed random variable with zero mean, $\sigmain N(0, 1)$.   Thus, our final data model is
\begin{equation}
\label{eq:dimming_empirical_law}
    \logten(\Lhi) = a + b \, \logten(\Llow) + \logten (C) + \sigmain N(0, 1).
\end{equation}
Here, $ C $'s distribution is given by $ f_C $ in \eref{eq:pdf_c} and included only when \mixing is assumed.
The resulting probability distribution for the total scatter $ S = \logten (C) + \sigmain N(0, 1) $ is the convolution between $ f_C $ and the Guassian distribution:
\begin{equation}
\label{eq:likelihood_alptheory}
    L_S \, ( s; p_0 ) \, = \,
    \frac{1}{\sqrt{2\pi}\sigmain}
    \int_0^1 \text{d}c \,
    \exp \left(-\frac{(s - \logten(C))^2}{2\sigmain^2}\right) \:
    f_C \, (c; p_0) \;.
\end{equation}

To determine whether \mixing actually occurred, we compare two different models: the {\it\Gaussmodel}, where the scatter from the empirical relation is simply Gaussian, and the {\it\ALPmodel}, where the dimming due to \mixing is super-imposed on the Gaussian scatter. Note that $\sigmain$ of the Gaussian scatter is empirically determined, and is normalised to match the observed scatter.  In the mixing model, it is assumed there is less intrinsic scatter, in order to keep the total scatter constant.  In fact, there is a minimum amount of scatter predicted by the \mixing model, and if a probe were found with less scatter than this, we could rule out the possibility of strong mixing for that probe.  The typical scatter coming from mixing alone is $ \sigmamix = 0.2 $, but this varies with the degree of initial polarisation.

\begin{figure}[htbp]
   \myfloatalign
      \includegraphics[width=\imagewidth]{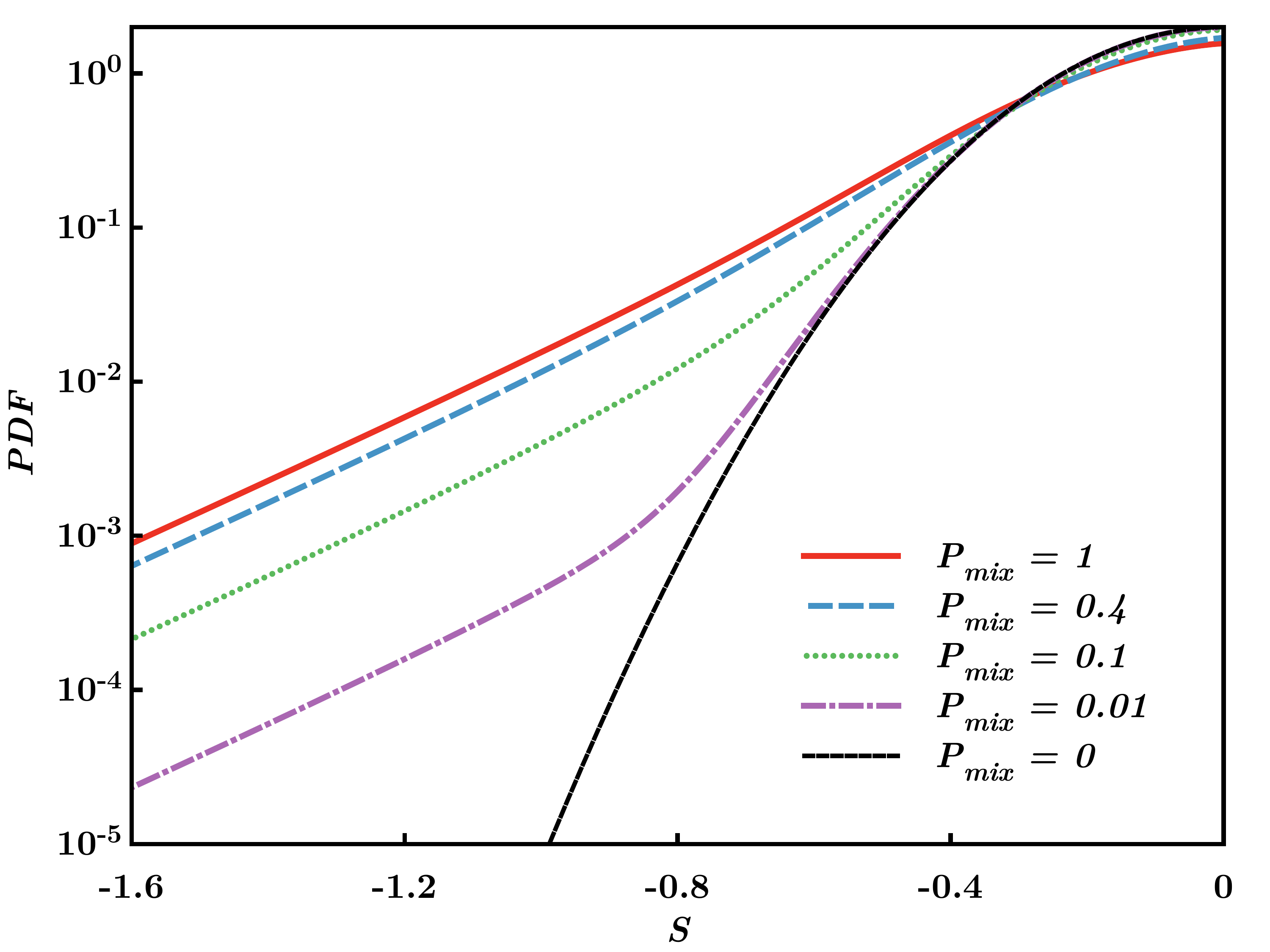}
   \caption{Probability distribution of the scatter in the \ALPmodel for various \Pmix values. We assumed $ p_0 = 0.1 $ and $ \sigmain = 0.2 $.
   For small \Pmix, the distribution is nearly Gaussian, apart from the low intensity tail.  Here and below we focus on the low end tail, where the relative deviations are largest.}
   \label{fig:pdf_s}
\end{figure}

Another factor to consider is the fraction of light paths which cross sufficiently magnetized regions to experience strong mixing.  There is no guarantee that any given source will live in a cluster environment, or that its light will pass through such an environment on its way to us.  Obviously, the less likely this is, the harder it will be to constrain the coupling to axion-like particles.  The final distribution will be a linear combination of the mixed distribution and the intrinsic distribution, weighted by the fraction of photon paths which experience mixing (\Pmix).  For small \Pmix the observed variance will be dominated by the intrinsic variance; the resulting distribution is very nearly Gaussian, but the mixing will significantly increase the low luminosity tail. The dependence of the likelihood from \Pmix is shown in \fref{fig:pdf_s}.

\subsection{Shot noise}
Finally, another potentially important contribution to the intrinsic scatter could be from shot noise in the numbers of \Xray or \gray photons detected.  This is of interest because it follows a Poisson distribution rather than Gaussian, and could have similar effects as the \mixing on the scatter.   The inferred \Xray luminosity is proportional to $ N $, while the intrinsic luminosity is proportional to \avg{N}, so that we have an additional contribution to the luminosity ratio which is $ \logten \NoverN $, where $ N $ is assumed to have a Poission distribution.

A very rough idea of the impact of the shot noise can be estimated by comparing its variance to that arising from the \mixing.  The \mixing variance  increases with the initial polarisation, but its minimum value is $ \var{A} = 0.033 $ when $ p_0 = 0 $.  The variance from shot noise depends on the number of photons observed, and the associated variances are shown in \tref{tab:shotnoise_variances}.   As can be seen, the variances are only comparable to \var{A} for $ N < 50 $, while typical surveys exceed this.    In the catalogs we consider below, all the objects have more than $ 50 $ net counts, with an average which is greater than $ 1000 $.  Thus, we do not expect naively that shot noise will be a major issue for these observations. On the other hand, objects in the catalogs considered by \citetalias{burrage:2009a} have photon counts as low as $ 10 $, with an average of $ 120 $; half of these sources have less than $ 50 $ net \Xray counts.

\begin{figure}[htbp]
 \myfloatalign
 \includegraphics[width=\imagewidth]{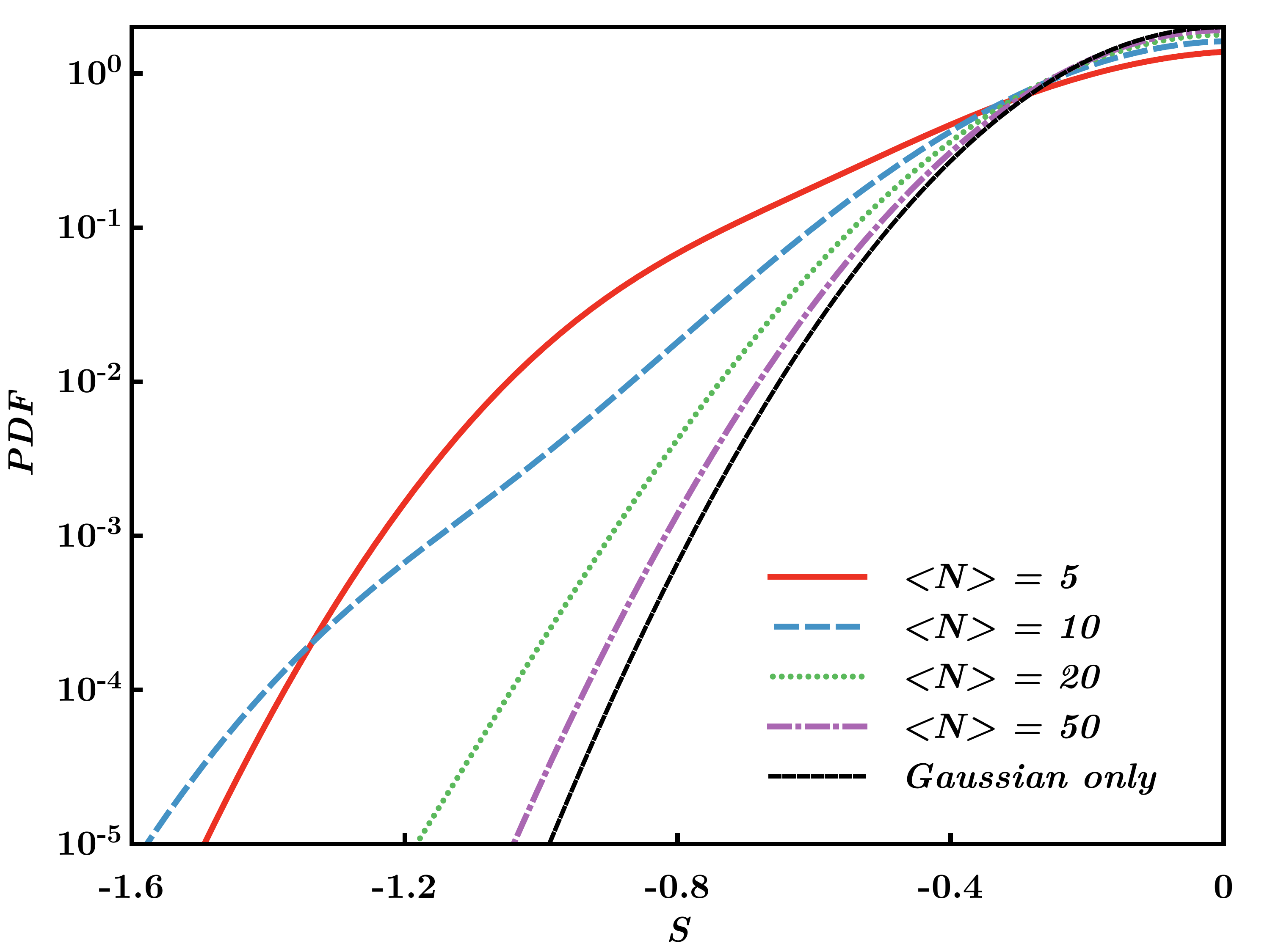}
 \caption{Probability distribution of the scatter taking into account shot-noise for various values of collected \Xray photons. We assumed $ \sigmain = 0.2 $ for the intrinsic Gaussian noise. Like the \ALPmodel, shot noise can increase the low luminosity tail.}  
 \label{fig:pdf_shotnoise}
\end{figure}

However, the contribution to the variance is only a rough proxy for the true effect, which is sensitive to the full distribution and in particular to the tail at low luminosities.  Also, if the fraction of sources which are strongly mixed is relatively low, then the shot noise can be a more significant issue.   See \fref{fig:pdf_shotnoise} for the probability distribution of the scatter in presence of shot-noise for various \avg{N} values. In \sref{sub:shotmodel_sim_results} we quantify this effect further for the statistical tests we consider.

\begin{table}[htbp]
    \begin{tabular}{ *{2}c }
        \avg{N} & \var{s} \\
       \hline \hline
        5 & 0.048 \\
        10 & 0.023 \\
        50 & 0.0038 \\
        100 & 0.0019 \\
        1000 & 0.00019 \\
        10000 & \sci{1.9}{-5} \\
    \end{tabular}
    \caption{Variance of the shot-noise contribution to the scatter for different average photon counts. The variance of the ALP contribution to the scatter is at least $ \var{A} = 0.033 $. In order to calculate \var{s} we did not consider the case in which $ N = 0 $ photons are collected, otherwise the variance would have been infinite. This is reasonable since $ N = 0 $ implies no measurement at all.}
\label{tab:shotnoise_variances}
\end{table}

\section{Statistical procedure}
\label{sec:stat_analysis}

\subsection{Bayes ratio}
We look at a number of statistical tests to examine whether the data support the \ALPmodel.   The Bayesian approach to comparing two models is to compare their Bayesian evidences; the evidence is the likelihood of the observed data given a model ($A$), integrated over the model parameters ($ {\bf{p}}_A $) and accounting for the prior distributions of the parameters:
\begin{equation}
{\cal P} ({\bf{D}}| A) = \int d^n {\bf{p}}_A \, {\cal P}({\bf{D}} | A,  {\bf{p}}_A ) \, {\cal P} ({\bf{p}}_A |A) \; .
\end{equation}
Here we wish to compare the evidences for the  \Gaussmodel and the \ALPmodel; for simplicity we fix the initial polarization $ p_0 $ and the mixing fraction, \Pmix.    The resulting models have the same parameters $ a $, $ b $ and \sigmain, and we assume the same flat prior distribution for these parameters.

If we make the further simplifying assumption that the shapes of the likelihoods of the models are the same (which must be established to first order by comparing the $ a $, $ b $ and $ \sigmain $ error bars of the models), the Bayes ratio of the models is simply given by the ratio of the peaks of the likelihoods:
\begin{equation}
\frac{{\cal P} ({\bf{D}}| ALP)} {{\cal P} ({\bf{D}}| G)} \simeq \frac{{\cal P}({\bf{D}} | ALP,  {\bf{\hat{p}}}_{ALP} )}{{\cal P}({\bf{D}} | G,  {\bf{\hat{p}}}_G )}.
\end{equation}
Here, ${\bf{\hat{p}}}$ are the parameters which maximise the likelihoods of the two models. This ratio is effectively equivalent to the $ r $ quantity used by \citetalias{burrage:2009a}:
\begin{equation}
\label{eq:likelihood_ratio}
    r\,(p_0, \Pmix) = 2 \, \ln
             \left[
                \frac{\hat{L}_{ALP}\,(p_0, \Pmix)}{\hat{L}_{G}}
             \right] \;.
\end{equation}
If $ r > 0 $, then the \ALPmodel is preferred over the \Gaussmodel; if $ r < 0 $ the opposite is true. The absolute value of $ r $ is distributed to a good approximation as a $ \chi^2 $ random variable with one degree of freedom. For example, $ \abs{r} = 9 $ corresponds to a $ \sim 3 \, \sigma $ preference for one model over the other.

Below we will be comparing the likelihoods of the various model parameters for different sets of data.  For the Gaussian model, the parameters include the amplitudes $ a $ and $ b $, and the intrinsic variance \sigmain.   For the mixing model, the parameters also include  $ p_0 $ and \Pmix.   The data are simply a collection of $ N $ observations of intensities of various sources in two different bands, $ X_i $ and $ Y_i $, which are taken to be independent.   The full resulting likelihood of the parameters given the data is:
\begin{widetext}
\begin{equation}
\label{eq:likelihood}
      L \, ( a, b, \sigmain; p_0, \Pmix ) \, = \,
      \prod_{i=1}^{N} \,
      \frac{1}{\sqrt{2\pi}\sigmain}
      \mathlineskip
      \left[
         \Pmix
         \int \text{d}c \, \exp \left(-\frac{z_i^2}{2\sigmain^2}\right) \: f_C \, (c; p_0)  \right.
         \mathlineskip
         \left. \!\! + \: (1-\Pmix) \exp \left(-\frac{s_i^2}{2\sigmain^2}\right) \:
      \right] \,,
\end{equation}
\end{widetext}
where $ s_i(a,b) =   \logten(\Lhi_i) - a - b \logten(\Llow_i) $,  $ z_i (a,b) = s_i(a,b) - \logten(\,C ) $ and we assumed \Pmix does not depend on redshift.

As discussed above, we then perform a Maximum Likelihood Estimate (MLE) of the parameters $ (a, b, \sigmain) $ for the \Gaussmodel ($ \Pmix = 0 $) and the \ALPmodel, leaving $ p_0 $ and \Pmix fixed. As a result, we obtain \emph{(i)} two sets of parameters: $ (\hat{a}, \hat{b}, \hat{\sigma}_{\text{in}})_{G} $ and $ (\hat{a}, \hat{b}, \hat{\sigma}_{\text{in}})_{ALP} $ and \emph{(ii)} the respective maximized likelihoods: $ \hat{L}_{G} $ and $ \hat{L}_{ALP}\,(p_0, \Pmix) $.

\subsection{Goodness-of-fit tests}
The Bayes ratio is the best means of comparing two models, but it does not examine whether either model provides a good fit to the data.   \BDS also looked at Bootstrap \citep{press:2002a} resamplings of the data sets, and showed plots of the variance versus skewness of the data (their so-called `fingerprints'), comparing these to what is expected in the Gaussian and ALP models.   In \sref{sec:bootstrap_bds} we shall reproduce their analysis.  In particular, we will show that much of the structure in these fingerprints arises from resampling of a few outliers multiple times and the resulting significance of such plots is hard to quantify.

Instead, we focus on the 1-D cumulative distributions of the scatter around the mean behavior which contain all of the relevant information.  For this kind of unbinned data, a standard goodness-of-fit statistic is the Kolmogorov-Smirnov (KS) test, which looks at the maximum difference of the cumulative distributions.   As we are fitting for parameters of the distribution, we simulate the process to see how often the observed KS statistic occurs in the two models.

As we shall see, statistics like the Bayes ratio are dominated by a few sources where the \Xray intensity is much lower than expected.   The KS test is not greatly sensitive to the tails of the distribution, so we also examine some related statistical tests, the Kuiper test and the Anderson-Darling (AD) statistic.  (A description of these can be found in \citet{press:2002a}.)  Briefly, the Kuiper statistic is the sum of the largest positive and negative difference in the observed and theoretical cumulative distributions, while the AD statistic is a renormalised version which gives more importance to the tails of the distribution.   Again, these tests are calibrated using simulations of the full process.

As is evident in the cumulative plots below, while the presence of the outliers can strongly favor the ALP distribution over the Gaussian, often many more outliers are seen than is expected by either model.   This most likely suggests that neither model is correct and that another explanation could be required for the low X-ray luminosity of some sources.  One strong candidate is that the soft X-rays are sometimes strongly absorbed.

\section{Data analysis}
\subsection{Previous analyses}
\label{sec:previous_analyses}
To discriminate between the \Gaussmodel and the \ALPmodel, \citetalias{burrage:2009a} analysed several classes of astrophysical objects where an empirical law of the form in \eref{eq:plain_empirical_law} is valid. Examples of such objects are Blazars, \gray bursts and Active Galactic Nuclei (AGN).  They obtained relevant results using a well-known correlation between the \ang{2500} and \kev{2} monochromatic luminosities of AGNs \citep{zamorani:1981a, avni:1982a, vignali:2003a, strateva:2005a, steffen:2006a, just:2007a, gibson:2008a, young:2009b}:
\begin{equation}
    \logten(\Lang{2500}) = a + b \logten(\Lkev{2}) .
\end{equation}

\citetalias{burrage:2009a} took into consideration \nagnBDSa optically selected AGN with redshifts less than $ 2.7 $ taken from \citet{steffen:2006a}. Of these, $ 32 $ are from the COMBO-17 survey \citep{wolf:2004a} and $ 45 $ from the Bright Quasar Survey (BQS) \citep{schmidt:1983a}. These two sets are matched with \Xray measurements coming respectively from the Extended Chandra Deep Field-South survey \citep{lehmer:2005a} and the \ROSAT experiment \citep{voges:1999a,pfeffermann:1987a}. We will refer to this set of AGNs as the \BDSa catalog. By analysing it and assuming $ \Pmix = 1 $, \citetalias{burrage:2009a} obtained:
\begin{equation*}
    r(p_0, \Pmix = 1) \simeq 14
\end{equation*}
for $ 0 \leq p_0 \lesssim 0.4 $ and $ r(p_0) \gtrsim 11 $ for $ p_0 > 0.4 $. This corresponds respectively to a $ 3.7\,\sigma $ and $ 3.3\, \sigma $ evidence in favor of the \ALPmodel.

\citet{burrage:2009c} recently extended their sample to \nagnBDSb AGNs by including $ 126 $ more AGNs with redshifts less than $ 3.8 $ from \citet{strateva:2005a}, optically selected from the Sloan Digital Sky Survey \citep{york:2000a} and matched for the most part with \ROSAT \Xray data \citep{voges:1999a,pfeffermann:1987a}. The net \Xray counts are in the range 10 -- 1500 with an average of $ 120 $ counts and $ 13\% $ of AGNs below $ 20 $ counts. We will refer to this set of AGNs as the \BDSb catalog. Using this catalog and assuming $ \Pmix = 1 $, they obtained the following result:
\begin{equation*}
    r(p_0 \lesssim 0.5, \Pmix = 1) \simeq 25 \;,
\end{equation*}
which corresponds to a 5$\sigma$ evidence in favor of the \ALPmodel.

\begin{figure*}[htbp]
    \myfloatalign
        \includegraphics[width=0.4\linewidth]{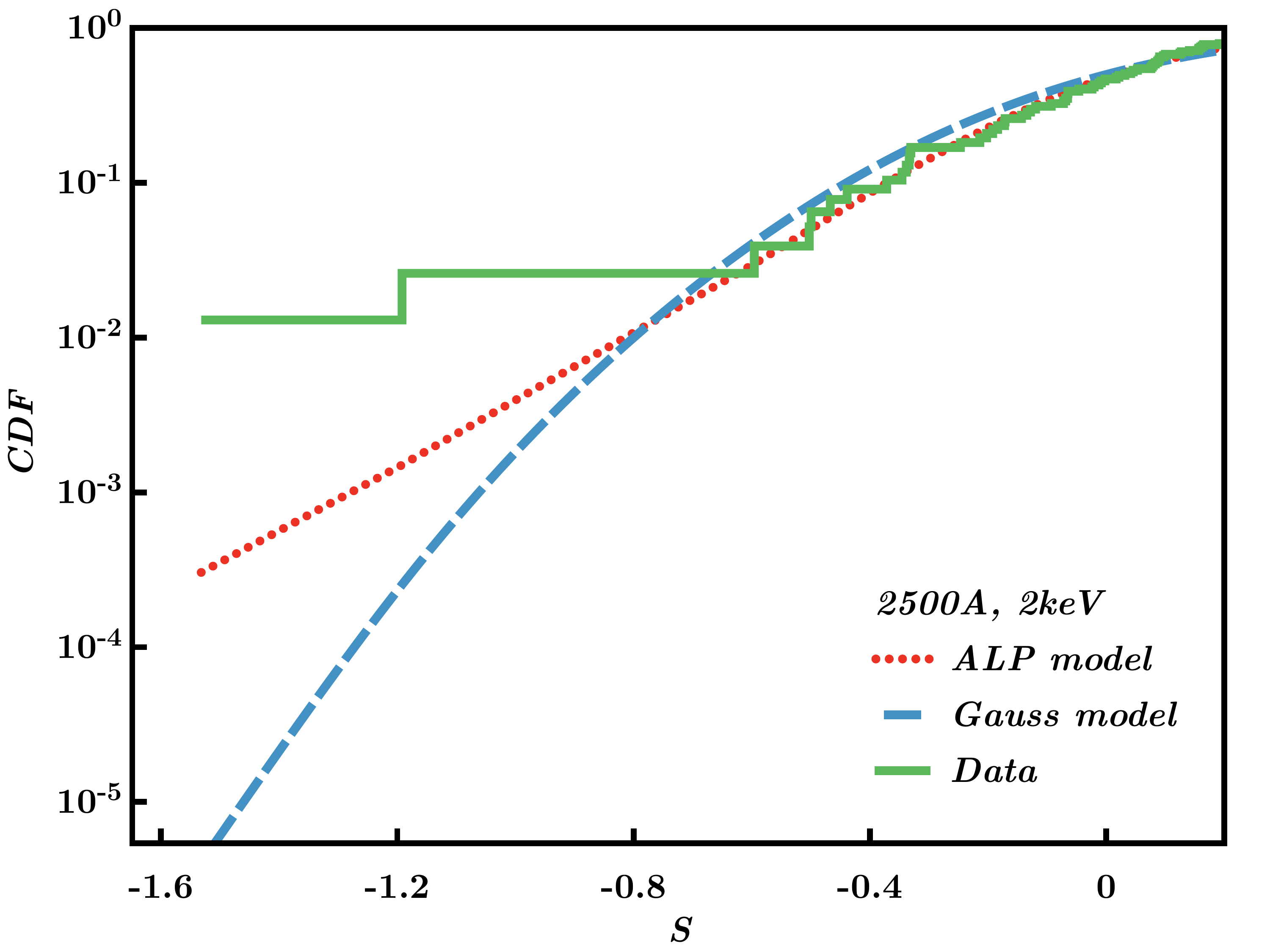}
        \includegraphics[width=0.4\linewidth]{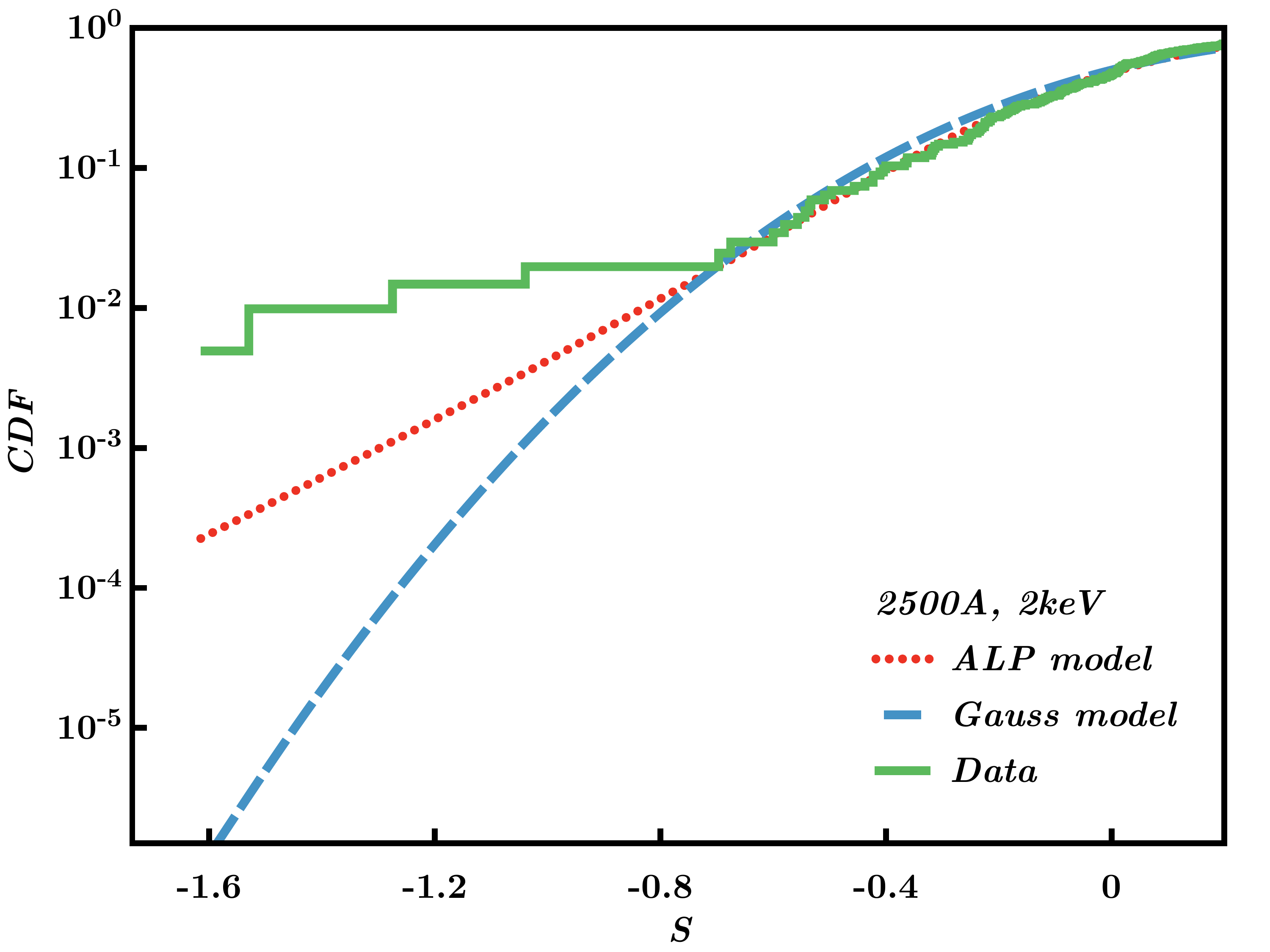}
    \caption{Cumulative Distribution Functions of the scatter for the \BDSa catalog (left panel) and \BDSb (right panel). Each empirical \CDF is plotted along with the \ALPmodel and \Gaussmodel theoretical \CDFs. We assumed $ p_0 = 0.1 $ and $ \Pmix = 1 $.}
    \label{fig:cdf_bds}
\end{figure*}

Below, we will quote only results where $ p_0 = 0.1 $ since \emph{(i)} this is the amount of linear polarisation predicted for AGN light \citep{bao:1997a} at $ E = \unit[2]{keV} $ and \emph{(ii)} the likelihood is almost insensitive to changes in $ p_0 $.

\subsubsection*{The problem of \Xray absorption}
\label{sub:problem_xray_absorption}
In addition to photon-ALP mixing, there are many other possible sources of scatter in the intrinsic relation between an AGN's optical or UV luminosity and its X-ray luminosity.  The X-rays are thought to arise from the hot coronal gas, while the lower energy photons are believed to radiate from the AGN accretion disk.  X-ray emission can be enhanced if there are jets, or suppressed if the coronae are absent or disrupted, or if there is significant absorption, which might occur as a result of outflows \citep{gibson:2008a}.
In addition, variation may occur because the time dependent X-ray and optical luminosities are measured at different epochs, though recent data have suggested that most of the scatter remains even for coeval observations \citep{vagnetti:2010a}.

Here we focus primarily on the effects of mixing, but if evidence suggests that there may be mixing, other sources of scatter must also be considered to explain the data.   The \rtest employed by \citetalias{burrage:2009a} is a simple likelihood ratio test.   The significant preference shown by the \BDSa and \BDSb catalogs for the \ALPmodel over the \Gaussmodel only demonstrates the relative fits of the models, but does not imply that either is actually a good fit to the data.   One way to evaluate the goodness of fit is to look at the Cumulative Distribution Functions (\CDF).   In \fref{fig:cdf_bds} we plot the \CDF of the scatter according to the two models against the empirical \CDF coming from these catalogs.    For both AGN sets, neither of the theory curves manages to reproduce the features of the scatter, though the \Gaussmodel is a much worse fit.   The empirical \CDF is much larger than the model \CDFs at the low end of the scatter axis,  meaning that the catalogs contain several objects with an \Xray to UV/optical luminosity which is much lower than what is likely via the \Xray dimming due to \mixing.

This effect can be naturally explained if we assume that \Xray light from these AGNs has been absorbed.  \citet{steffen:2006a} attempted to remove this possible contamination by excluding AGNs with flat \Xray spectra according to the effective \Xray power-law photon index $ \Gamma $ (some details on this procedure can be found in \sref{subs:highgammaset}); however, this was only possible for the Chandra subset, leaving some potentially contaminated AGN which can dominate the statistics.   Unfortunately, the $ r $ statistic is easily dominated by outliers, so that even a small contamination can significantly bias the result.

To emphasise the effect of the unaccounted for \Xray absorption, we exclude from the \BDSa and \BDSb catalogs one $ z = 0.067 $ AGN  (PG2214+139, also known as MKN 304), which is known to be heavily obscured in the \Xray \citep{piconcelli:2004a, piconcelli:2005a}.  This source has a very flat \Xray spectrum dominated by strong absorption features due to ionized gas.  The gas is well modeled by a two-component warm gas, which yields one of the highest column density in XMM-Newton and Chandra.
PG2214+139 is the biggest outlier in both datasets and has a scatter $ \sim 4.5 $ standard deviations below the average.  It is actually visible in \fref{fig:cdf_bds} as the leftmost point in both plots.  By repeating the \rtest without PG2214+139, we obtain a drop in the \rstat of $ 64\% $ for the \BDSa  catalog and of $ 36\% $ for the \BDSb catalog.

Another AGN in the \BDSb catalog that deserves special attention is TGN336Z208, known in SDSS as SDSSJ134351.12+000434, which has been spectroscopically classified \citep{georgakakis:2004a} as a Broad Absorption Line (BAL) AGN. It is thought that BALs are AGNs viewed through the non-spherical wind that surrounds the accretion disk of the supermassive black hole powering AGN emission (see \citet{murray:1995a}). The \Xray absorption resulting from this obscuration renders BAL AGNs unsuited to study the intrinsic correlation between their UV and X-ray emissions \citep{brandt:2000a, gallagher:2002a}.%
\footnote{The selection criteria in forming the \BDSb catalog \citep{strateva:2005a} included removal of BALs according to the properties of the \ion{C}{iv} and \ion{Mg}{ii} absorption lines. However, these can be seen in SDSS spectra only for sources with $ z > 1.55 $ and $ z > 0.45 $ respectively, while TGN336Z208 has $ z = 0.0736 $.}
Moreover, \citet{strateva:2005a} only report $ 24 $ net \Xray counts for TGN336Z208, a number which is too low to permit model fitting on the spectrum.   By removing both SDSSJ134351.12+000434.8 and PG2214+139 from the \BDSb catalog, the $ r $ statistic drops from $ r \simeq 25 $ to $ r \simeq 7 $, a $ \sim 70\% $ decrease.

Inevitably, our focus was drawn to these particular AGN because they were less X-ray bright and dominate the Bayes ratio test; this means that some care must be taken to treat all the data consistently to avoid a posterior bias which would occur from arguing selectively to omit those AGN with the highest scatter.  However, we have shown that there are independent reasons, based on their spectroscopic properties, for excluding these from the sample.  In any case, these results emphasise the great sensitivity of this test to outliers and the need to ensure a sample free from absorption.

\subsubsection*{Bootstrap resamplings}
\label{sec:bootstrap_bds}
\begin{figure}[htbp]
    \myfloatalign
    \subfloat[]{
        \label{fig:fingerprints_bdsa}
        \includegraphics[width=0.32\linewidth]{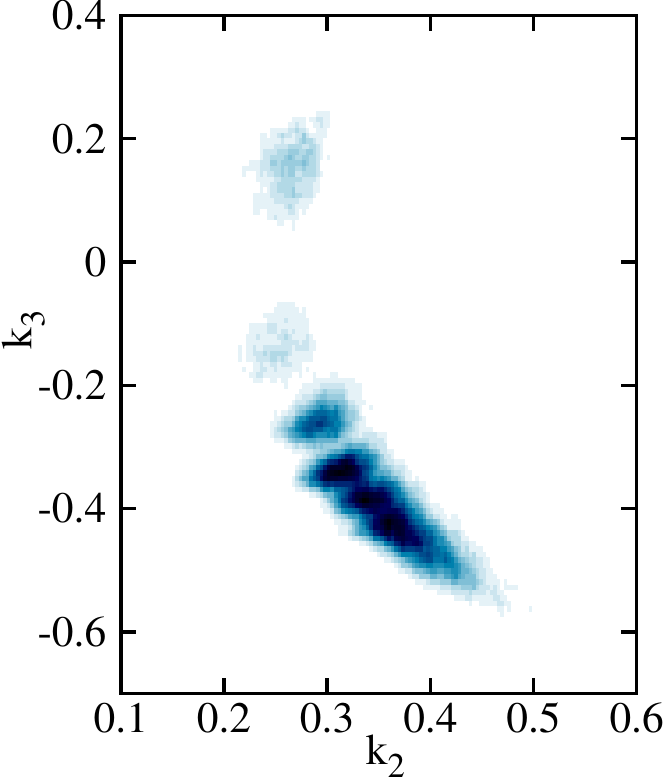}
    }
    \subfloat[]{
        \label{fig:fingerprints_bdsa_typical}
        \includegraphics[width=0.32\linewidth]{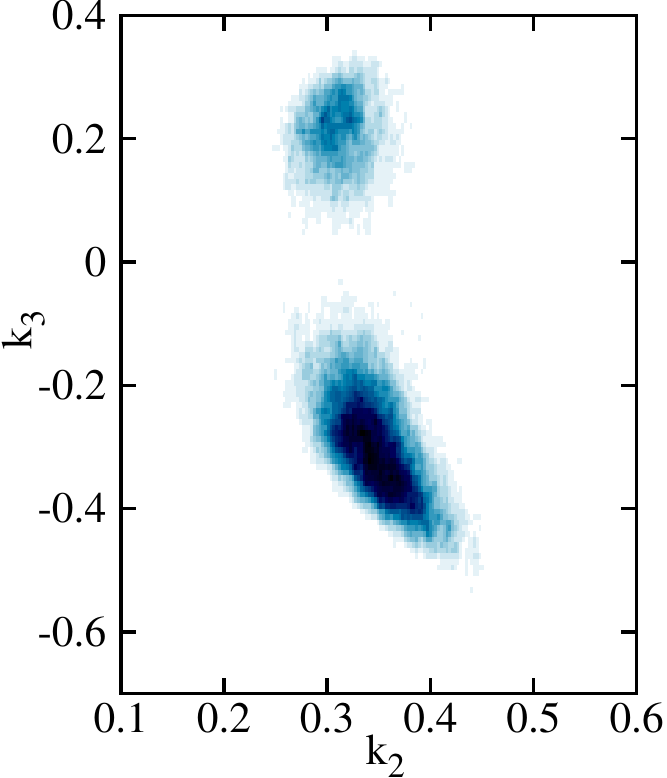}
    }
    \subfloat[]{
        \label{fig:fingerprints_bdsa_outlier}
        \includegraphics[width=0.32\linewidth]{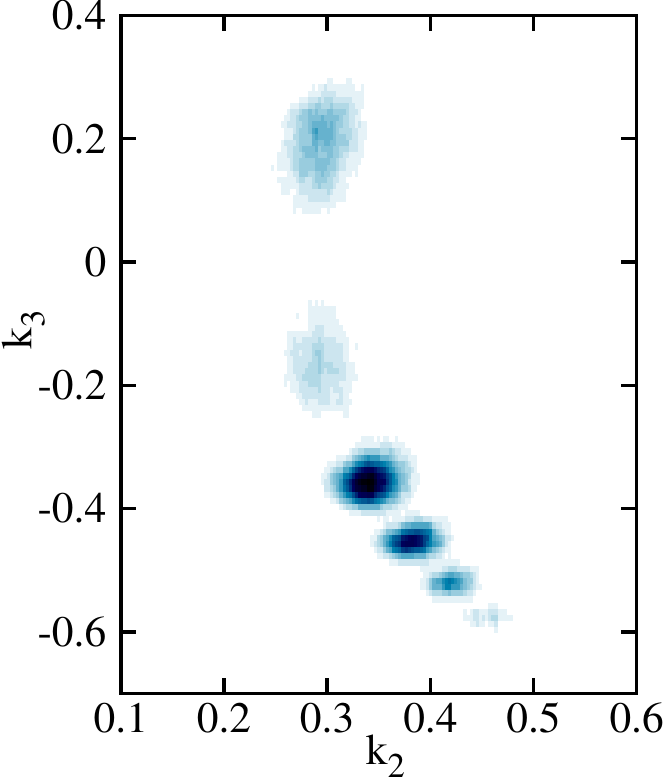}
    }
    \caption{Fingerprint plots of variance versus skewness.  Each point represents a Bootstrap resampling of \nagnBDSa data points \subref{fig:fingerprints_bdsa} from the \BDSa catalog of AGNs, \subref{fig:fingerprints_bdsa_typical} simulated with the \ALPmodel and \subref{fig:fingerprints_bdsa_outlier} simulated with the \ALPmodel, but where there happens to be a large outlier.}
\end{figure}

As further evidence of the \ALPmodel, \BDS introduced the concept of `fingerprints', which are plots based on Bootstrap resamplings of the AGN data set.   They created \nresamp data sets derived from the \BDSa catalog, each of the same size as the original catalog, by sampling it with replacement.   For each of these resampled data sets, they calculated the moments of the distributions, defined as      
\begin{equation*}
  k_m = \left(\frac{1}{N} \sum_{i=1}^N {S_i}^2\right)^{1/m} \;,
\end{equation*}
where $ N $ is the number of AGNs in the sample and $ S_i $ is the scatter of the \emph{i}-th AGN.   The fingerprints are scatter plots of the moments (e.g. the variance, $k_2$, versus the skewness, $k_3$) for all of the resampled data sets.  We have reproduced this 
analysis and it can be seen in \fref{fig:fingerprints_bdsa}.  

\BDS showed that there were similarities between the scatter plots generated from the data and those generated from a sample simulated with the \ALPmodel which were not seen in the simple Gaussian case.  Example simulations can be seen in  Figures \ref{fig:fingerprints_bdsa_typical} and  \ref{fig:fingerprints_bdsa_outlier}.  The data and the simulations both share a similar tail to large variance and negative skewness, which are rare in Gaussian simulations.  In addition, the data and  \fref{fig:fingerprints_bdsa_outlier} also have a similar substructure in this tail.

However, these fingerprints are similarly sensitive to outliers in the data set which is being resampled.  For example, the substructure apparent in Figure \ref{fig:fingerprints_bdsa_outlier} is due to a single outlier more than $ 4\sigma $ away from the mean.   In some resamplings, this outlier does not appear, resulting in two islands of low variance and small skewness (which can be positive or negative); in other resamplings it can appear once, resulting in another island with higher variance and more negative skewness.   The outlier can be resampled multiple times, and islands of substructure arise associated with the outlier appearing two, three or even four times.   The actual data is similar, but there is one large outlier and one moderate outlier which provide somewhat finer structure.

Substructure, since it results from outliers, is more likely to arise from the \ALPmodel than in the Gaussian model because of its higher tail (\fref{fig:pdf_s}.)  However, even in the \ALPmodel such substructure is not common in resamplings based on typical simulations; a more typical result is shown in  \fref{fig:fingerprints_bdsa_typical}.  Even in the data, when the data sets get bigger, the impact of a single data point is smaller and the substructure is less apparent.  (For example, significantly less substructure was seen in the \BDSb catalog.) 

The tail to large variance and negative skewness does remain in the data and the simulations of the \ALPmodel and is absent from the Gaussian models.  This reflects the fact that both distributions are skewed, having larger tails on the side with lower luminosities.   However, the similarities are qualitative, and the fingerprints have not been used to quantify the size of the tails, where the data and the \ALPmodel differ significantly (as was seen in \fref{fig:cdf_bds}.)

%

\subsection{Defining our AGN samples}
Here we take advantage of a new AGN sample, using more than $ 300 $ AGNs from the Fifth Data Release Sloan Digital Sky Survey/Xmm-Newton Quasar Survey (\citet{young:2009b, young:2009c}).  This sample contains three optical monochromatic luminosities and six \Xray monochromatic luminosities.  The multi-band data allow us to compute  the $ r $ statistic for $ 18 $  different combinations of optical/UV and \Xray monochromatic luminosities.  In addition, spectral fits of each AGN aid in excluding \Xray obscured AGNs from our sample.  The harder \Xray luminosities available in this sample should also be less subject to absorption. 

\subsubsection{The \sdssxmm catalog}
\label{subs:fullset}
The \sdssxmm catalog contains $ 792 $ AGNs. We first select \nagnFull AGNs by applying the same cuts performed in \citet{young:2009b}:
\begin{enumerate}
    \item we exclude all Radio Loud Quasars;
    \item we exclude all Broad Absorption Line Quasars;
    \item we consider only AGNs with \Xray detections characterised by a signal-to-noise ratio (S/N) greater than 6. The objects we are left with have net \Xray counts in the range 51 -- 39300, with an average of $ 1336 $ counts (so that shot noise is minimal). This allows for spectral fits to be made over the $ \unit[0.5 - 10]{keV} $ band for each source;
    \item we select AGNs whose preferred \Xray spectral fit is a single power-law (SPL) with no intrinsic absorption;
    \item between the remaining AGNs, we choose only those with a good spectral fit, that is with a reduced $ \chi^2 $ statistic (\ie $ \chi^2 / \text{d.o.f.} $) smaller than $ 1.2 $.
\end{enumerate}
The resulting set of AGNs lies in the redshift range 0.1 -- 4.4. We will refer to it as the \Full catalog.

\subsubsection{Removing obscured AGNs}
\label{subs:highgammaset}
We consider also a further cut on the effective \Xray power-law photon index $ \Gamma $ (column $ 10 $ in Tab.\ 2 of \citet{young:2009c}).  The photon spectral index is defined as $ \Gamma = -\alpha + 1 \;,$ where $ - \alpha $ is the exponent obtained by fitting the \Xray part of the AGN spectrum with a single power-law model.  A low $ \Gamma $ value implies a flat spectrum, that is a spectrum where the soft \Xray component (\ie the $ \unit[0.5 - 2]{keV} $ band) is much weaker than the hard \Xray component (\ie the $ \unit[2 - 10]{keV} $ band).  There is strong evidence \citep{brandt:2000a, gallagher:2002a, piconcelli:2005a, gibson:2008a} that \Xray absorption is the primary cause of soft \Xray weakness.  Intuitively, this happens because the low energy \Xrays are more easily absorbed than high energy ones.

Following \citet{steffen:2006a}, we select only those AGNs where $ \Gamma > 1.6 \;, $ thus reducing our sample to \nagnHigh AGNs.  We will refer to this as the High-$\Gamma$ catalog.
One can further reduce the impact of \Xray absorption by imposing cuts on the signal-to-noise ratio; it is possible, for example, to form AGN catalogs where $ S/N > 10 $ or $ S/N > 20 $.  We statistically analysed these samples as well and we ended up with results very similar to those obtained from the High-$\Gamma$ catalog. Moreover, we applied cuts on the Galactic column density in the direction of the sources and did not find any significant correlation with the \rstat.

\subsubsection{Multi-wavelength data}
\label{subs:multi-wave}
A significant advantage of the sample of \citeauthor{young:2009b} is its many frequency bands.
This consists in the rest-frame monochromatic luminosities at the following frequencies: \ang{1500}, \ang{2500}, \ang{5000} for the optical/UV side and \kev{1}, \kev{1.5}, \kev{2}, \kev{4}, \kev{7}, \kev{10} for the \Xray side. By applying the statistical analysis outlined in \sref{sec:stat_analysis} to multi-wavelength data, we can check whether the $ r $ statistic varies as the \Xray frequency goes from the soft to the hard side of the spectrum. The \ALPmodel does not predict any variation since, as long as we are in the frequency-independent limit (see \sref{sec:introduction}), the oscillation probability in \eref{eq:alp_oscillation_prob} is insensitive to the photon energy. On the other hand, soft \Xrays are more easily absorbed than hard \Xrays. It is therefore clear that a multi-wavelength analysis can help discriminating between photon dimming due to \mixing and \Xray absorption.

\subsection{Results}
\label{sec:rtest_results}
\begin{figure}[htbp]
    \myfloatalign
        \includegraphics[width=\imagewidth]{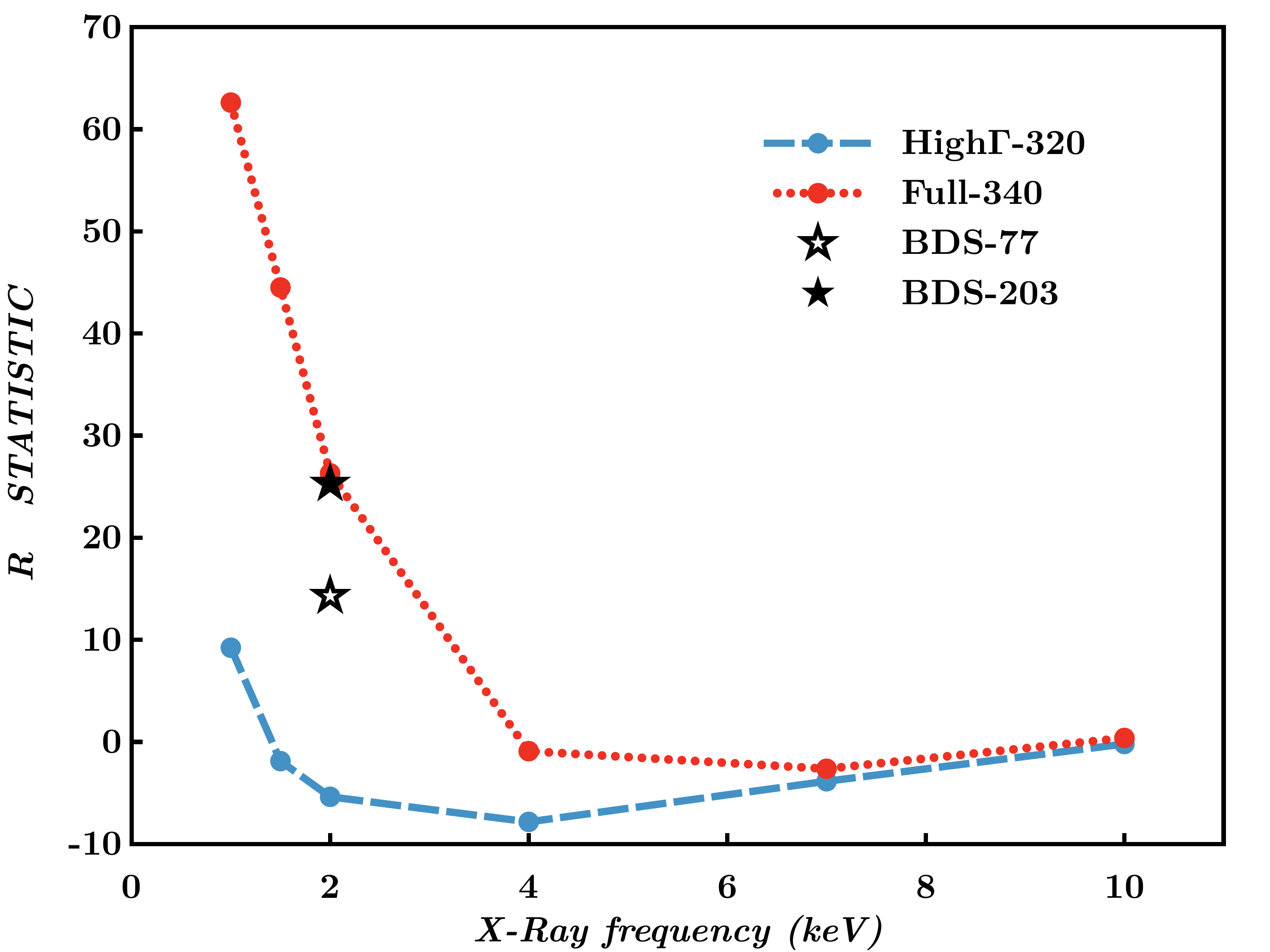}
    \caption{Value of the \rstat for the various wavelengths and catalogs (\Pmix =1).  The evidence for ALPs is significantly lower in those bands where the effects of absorption are expected to be less.}
    \label{fig:r_plot}
\end{figure}

In \tref{tab:r_results} we report the results we obtained by analysing the Full and High-$\Gamma$ catalogs using the method outlined in \sref{sec:stat_analysis} at the various wavelengths and for four \Pmix values. \fref{fig:r_plot} shows the same results in a $ r $-vs-\Xray energy plot. In the following discussion we shall always assume an optical/UV wavelength of \ang{2500} since the dependence on optical wavelength is barely noticeable.

\subsubsection{\Full catalog}
The outcome of the \rtest significantly depends on the analysed \Xray frequency. Regardless of \Pmix, the \ALPmodel is strongly preferred in the soft \Xray band (\unit[1]{keV}, \unit[1.5]{keV} and \unit[2]{keV}). For example, for $ E = \unit[1]{keV} $ and $ \Pmix = 1 $ we have $ r \simeq 63 $, a $ \sim 8\sigma $ preference for the \ALPmodel. However, as soon as we move to the hard \Xray bands (\unit[4]{keV}, \unit[7]{keV} and \unit[10]{keV}), $ r $ quickly approaches zero: neither model is preferred.  The energy dependence suggests that \Xray absorption, rather than \mixing, represents the main contribution to the scatter at low energies.  Where the $r$ value is high, the \PDFs are not especially consistent with the \ALP model or the Gaussian model.  

\begin{table}[htbp]
    \begin{tabular}{l *{6}r }
        & \kev{1} & \kev{1.5} & \kev{2} & \kev{4} & \kev{7} & \kev{10}\\
        \hline \hline
        \multirow{2}{*}{\Pmix = 1.0}
                    & 63 & 45 & 26 & -0.91 & -2.6 & 0.39\\
                    & 9.2 & -1.9 & -5.4 & -7.8 & -3.9 & -0.20\\
        \hline
        \multirow{2}{*}{\Pmix = \pmixhigh}
                    & 62 & 43 & 25 & -1.3 & -2.8 & 0.28\\
                    & 8.6 & -2.8 & -6.2 & -8.2 & -4.0 & -0.34\\
        \hline
        \multirow{2}{*}{\Pmix = \pmixmiddle}
                    & 60 & 44 & 26 & 0.60 & -1.5 & 0.86\\
                    & 10 & 0.87 & -2.4 & -5.6 & -2.6 & 0.38\\
        \hline
        \multirow{2}{*}{\Pmix = \pmixlow}
                    & 42 & 32 & 19 & 2.4 & -0.12 & 0.66\\
                    & 9.2 & 4.7 & 2.6 & -0.90 & -0.50 & 0.55\\
    \end{tabular}
    \caption{Outcome of the $ r $-test applied to the \Full (upper line) and \HighG (lower line) catalogs at the various wavelengths and for different mixing fractions. Positive values imply that the \ALPmodel is preferred; negative values imply that the \Gaussmodel is preferred. We assumed $ p_0 = 0.1 $, but the results do not change greatly for different values.}
\label{tab:r_results}
\end{table}

\begin{figure*}[htbp]
    \myfloatalign
    \subfloat[\Full]{
        \label{fig:cdf_fullset}
        \includegraphics[width=\imagewidth]{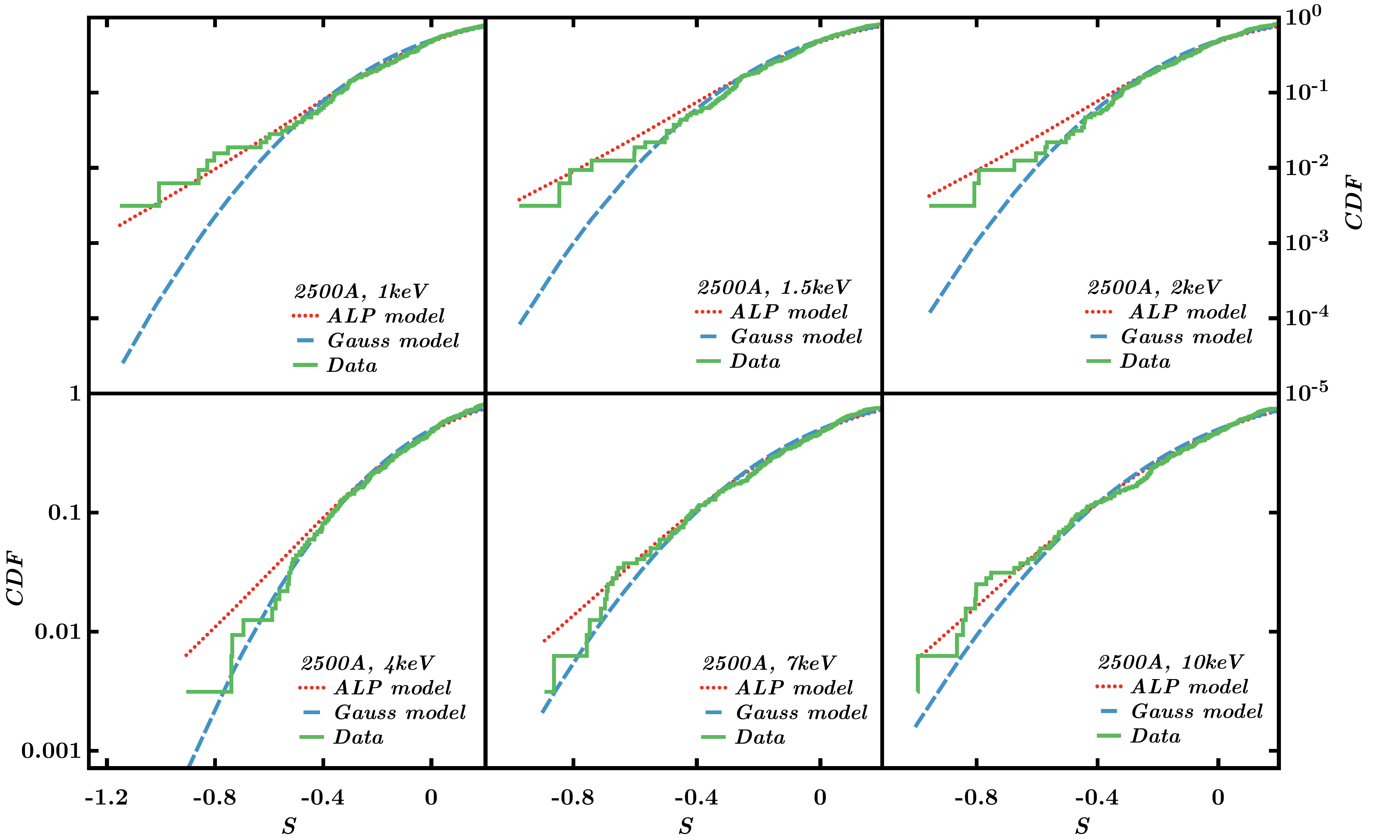}
    } \\
    \subfloat[\HighG]{
        \label{fig:cdf_highgammaset}
        \includegraphics[width=\imagewidth]{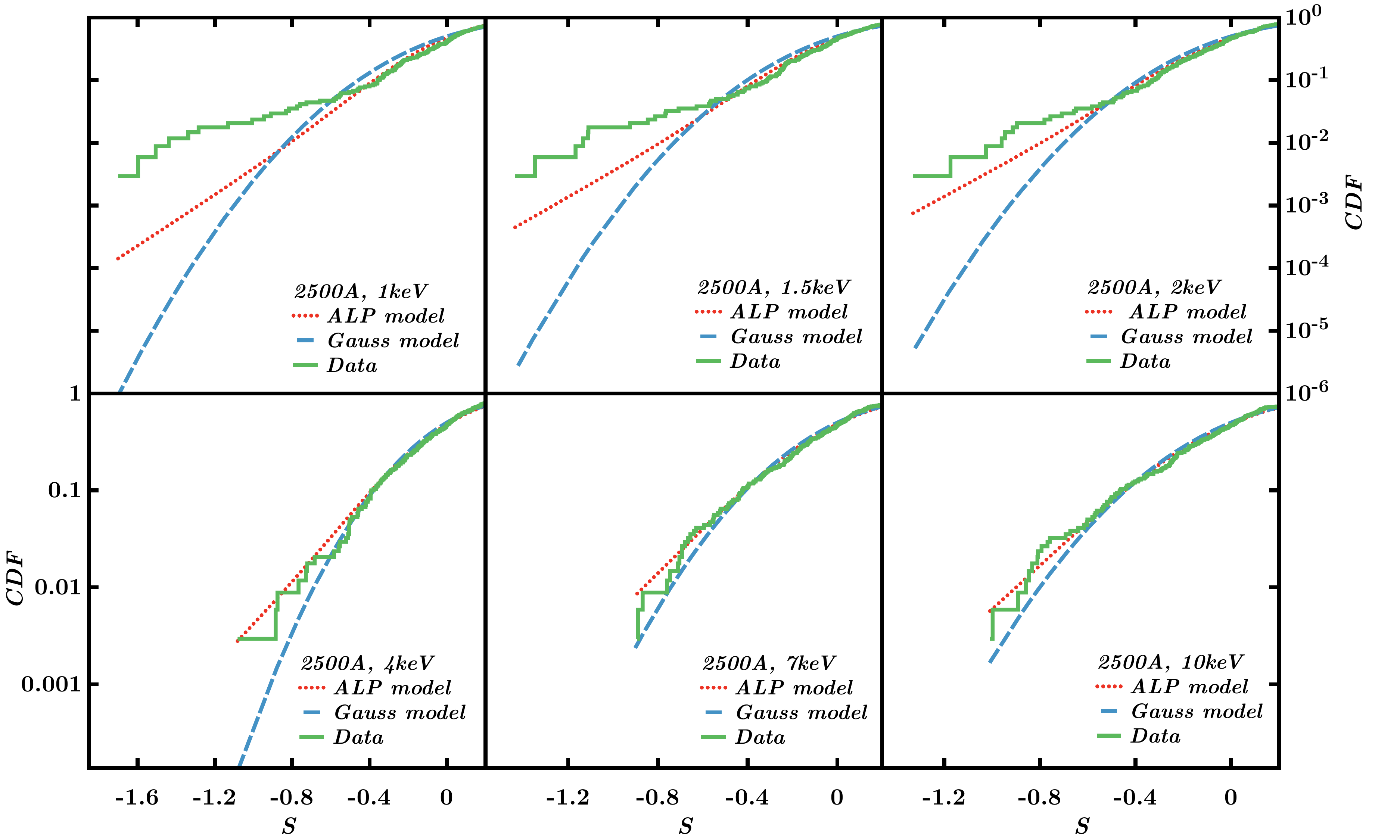}
    }
    \caption{Cumulative distribution functions of the scatter for the Full \subref{fig:cdf_fullset} and High-$\Gamma$ \subref{fig:cdf_highgammaset} catalogs. The empirical \CDF is plotted along with the \ALPmodel and \Gaussmodel theoretical \CDFs. For the \ALPmodel we assumed $ p_0 = 0.1 $ and $ \Pmix = 1 $. When $ \Pmix < 1 $, the data curve stays the same while the \CDF of the \ALPmodel tends to that of the \Gaussmodel.}
    \label{fig:cdf_sdssxmm}
\end{figure*}

In \fref{fig:cdf_fullset} we plot the Cumulative Distribution Functions (\CDF) for the \ALPmodel and the \Gaussmodel together with the empirical \CDF of the \Full catalog. In the soft \Xray band both the \ALPmodel and the \Gaussmodel fail to reproduce the features of the scatter. The preference for the \ALPmodel over the \Gaussmodel given by the \rtest at these energies is due to a very bad performance of the latter rather than to a good performance of the former. Moreover, both models systematically overestimate the \Xray luminosities: \Xray absorption, again, could be the culprit of this discrepancy.   In the hard \Xray band, where absorption is less likely to affect the photons, the empirical \CDF stays in between the two theoretical curves: no model is to be preferred.

\subsubsection{\HighG catalog}
The signal for the \ALPmodel dramatically drops upon removing the \nagnLowG AGNs which have lower photon spectral index. As shown in \tref{tab:r_results}, this is true for all the considered \Pmix values. These are explained looking at the High-$\Gamma$ \CDFs in \fref{fig:cdf_highgammaset}.

At the lowest energies, the distribution is very much like the expected ALP prediction, especially in the \kev{1} case.  However, we do not expect the \mixing signal to be frequency dependent, while we see the \Xray absorption is. This suggests also that the agreement with the ALP model is likely the accidental effect of a small amount of residual \Xray intrinsic absorption, rather than due to \mixing.

\section{Simulated data sets}
\label{sec:r_distribution}
It is useful to understand the sensitivity to the question of mixing which we expect to find given the catalogs we consider.   Using simulated samples drawn from the mixing distribution, we can measure the distribution of $ r $-values which would typically occur and then compare these to the $ r $-values obtained in \sref{sec:rtest_results}.   If we measure a significantly different value, this could indicate that there may be an alternative explanation of the signal.

The same simulated data sets can be used to derive the statistical significance of the goodness-of-fit tests introduced in \sref{sec:stat_analysis}, namely the Kolmogorov-Smirnov (KS), Kuiper and Anderson-Darling (AD) tests. We will do so in \sref{sec:Goodness-of-fit_tests}.

For the models we analysed in \sref{sec:model_assumptions} -- the \ALPmodel, the \Gaussmodel and the shot-noise model --~, the \PDF's are known and it is straightforward to produce samples consistent with each.  For each assumed scatter model, we adjust the parameters to match those of the observed samples; in particular we match the the total variance $ \sigmatot^2 $ of the scatter distribution for each sample, as well as the number $ \nagn $ of AGNs in each sample.  We consider the three combinations of \nagn and $ \sigmatot^2 $ corresponding to the \BDSb, \Full and \HighG catalogs. All values of \nagn and $ \sigmatot $ are shown in \tref{tab:nagn_sigmatot_catalogs}.

\begin{table}[htbp]
    \begin{tabular}{ l | c | l*{5}c }
		Catalog   & $ \nagn $    & \sigmakev{1} & \sigmakev{1.5} & \sigmakev{2} & \sigmakev{4} & \sigmakev{7} & \sigmakev{10}  \\
        \hline \hline
        \BDSb  & \nagnBDSb & & & 0.340 & & & \\
        \Full & \nagnFull  & 0.357 & 0.314 & 0.300 & 0.296 & 0.319 & 0.344 \\
        \HighG & \nagnHigh & 0.282 & 0.259 & 0.260 & 0.282 & 0.315 & 0.340
    \end{tabular}
    \caption{Combinations of $ \nagn $ and $ \sigmatot $ used to generate simulated scatter samples of the AGN catalogs.}
\label{tab:nagn_sigmatot_catalogs}
\end{table}

For each model and each dataset, we generate \nsimsrtest scatter samples, and apply the \rtest outlined in \sref{sec:stat_analysis} to each generated sample.    Histograms of these results are shown in the figures below. Each figure reports also the measured signal as a vertical black line and its statistical significance (\ie the value of the empirical \CDF) in the legend.

\subsection{\ALPmodel results}
\label{sub:alpmodel_sim_results}
In Figures \ref{fig:r_hist_bdsb_x2_alp}, \ref{fig:r_hist_full_x2_alp} and \ref{fig:r_hist_highg_x2_alp}, we show the $ r $-distributions obtained for the various catalogs for $ E = \kev{2} $. Before comparing them with the measured signals, let us point out some general properties of the \rtest which stem from these distributions.

First, the \rtest may not be sufficient to provide a conclusive preference for \mixing even if the latter is actually happening. Consider the expectation value of the \rstat for an AGN sample with $ \nagn = \nagnBDSb $ and $ \sigmatot = 0.34 $ (\ie the \BDSb catalog), in the best scenario of $ \Pmix = 1 $. This amounts to $ \avg{r} = 3.2 $, which corresponds to an evidence for the \ALPmodel of less than 2$\sigma$. Even if \mixing were happening, the \BDSb catalog would usually not be enough to provide a detection. The statistical significance attainable with the \rtest increases with \nagn and \Pmix and decreases with the intrinsic variance of the dataset $ \sigmain^2 $. For the \Full and \HighG catalogs, we have $ \avg{r} \sim 9 $ and $ \avg{r} \sim 21 $ which correspond to a preference for the \ALPmodel of 3$\sigma$ and 4.5$\sigma$ respectively.

When $ \Pmix < 1 $, we expect it to be harder to discriminate between the \ALPmodel and the \Gaussmodel and, therefore, the average of the \rstat should be smaller. The simulations confirm this behaviour.  For example, if $ \Pmix = \pmixmiddle $, then the average $ r $-values for the Full and High-$\Gamma$ data sets shrink to $ \avg{r} \sim 6.6 $ and $ \avg{r} \sim 13 $, that is a 2.6$\sigma$ and 3.6$\sigma$ preference respectively.

The figures also demonstrate another issue: the probability of a false-negative result, \ie the \rtest yielding a preference for the \Gaussmodel when the scatter comes from the \ALPmodel, is not negligible even when $ \Pmix = 1 $. We define the false-negative probability as the value of the empirical \CDF at $ r = 0 $. For the \BDSb catalog, this amounts to $ 0.27 $: there is a $ 27\% $ probability that the \rtest gives a false-negative result when applied to a dataset similar to the \BDSb catalog. The false-negative probability for the \Full and \HighG catalogs is smaller since both have a smaller scatter variance than \BDSb, but is still significant: $ 11\% $ and $ 4\% $ respectively.

Clearly the limiting factors are the number \nagn of AGNs and the intrinsic variance $ \sigmain^2 $ of the scatter of the empirical relation taken into consideration.  The contribution to the scatter from mixing is fixed (for a given \Pmix), so if the total scatter is larger, the greater will be the relative contribution from the intrinsic scatter and the harder it is to distinguish \mixing.
Statistically significant results require using datasets larger than those considered in this paper. For example, $ 500 $ AGNs with $ \sigmatot = 0.26 $, in the best scenario of $ \Pmix = 1 $, lead to an average evidence 5.7$\sigma$, while $ 1000 $ objects yield an average evidence of 7.7$\sigma$.

\subsubsection*{Comparison with the measured signal}
\begin{figure}[htbp]
    \myfloatalign \includegraphics[width=\imagewidth]{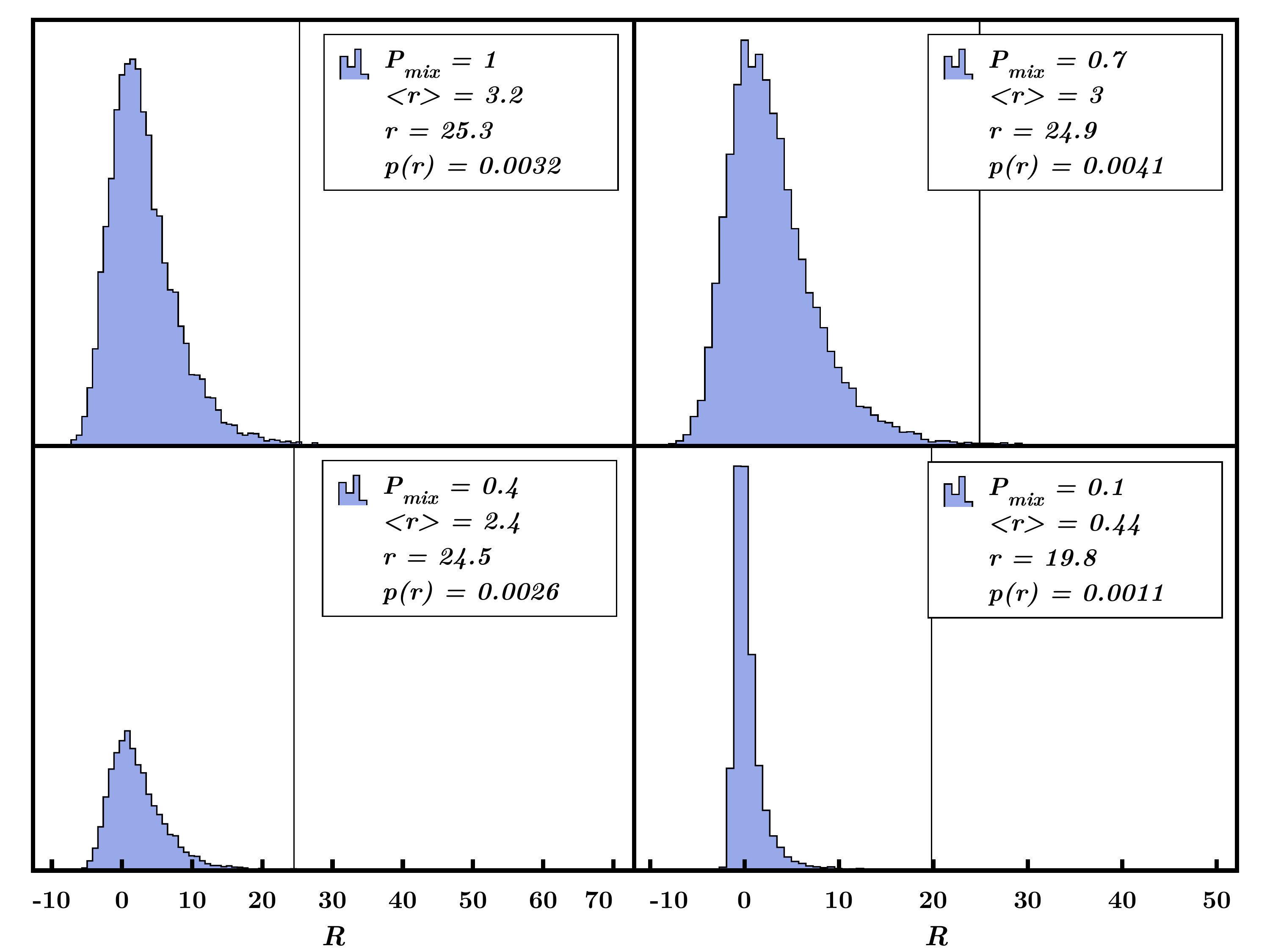}
    \caption{\rdist for the \BDSb catalog at $ E = \kev{2} $ when the scatter is distributed according to the \ALPmodel. The measured signal, $ r $, is represented by the vertical black line, while its statistical significance is quoted in the legend as $ p(r) $.
	}
    \label{fig:r_hist_bdsb_x2_alp}
\end{figure}

The measured $ r $-value for the \BDSb catalog is much larger than what would typically occur when applying the \rtest to a scatter sample distributed according to the \ALPmodel.   As can be seen by \fref{fig:r_hist_bdsb_x2_alp}, regardless of the considered \Pmix, we find the probability of the observed signal to be smaller than $ 0.4\% $.   We obtain a similar result for the \BDSa catalog.

\begin{figure}[htbp]
    \myfloatalign \includegraphics[width=\imagewidth]{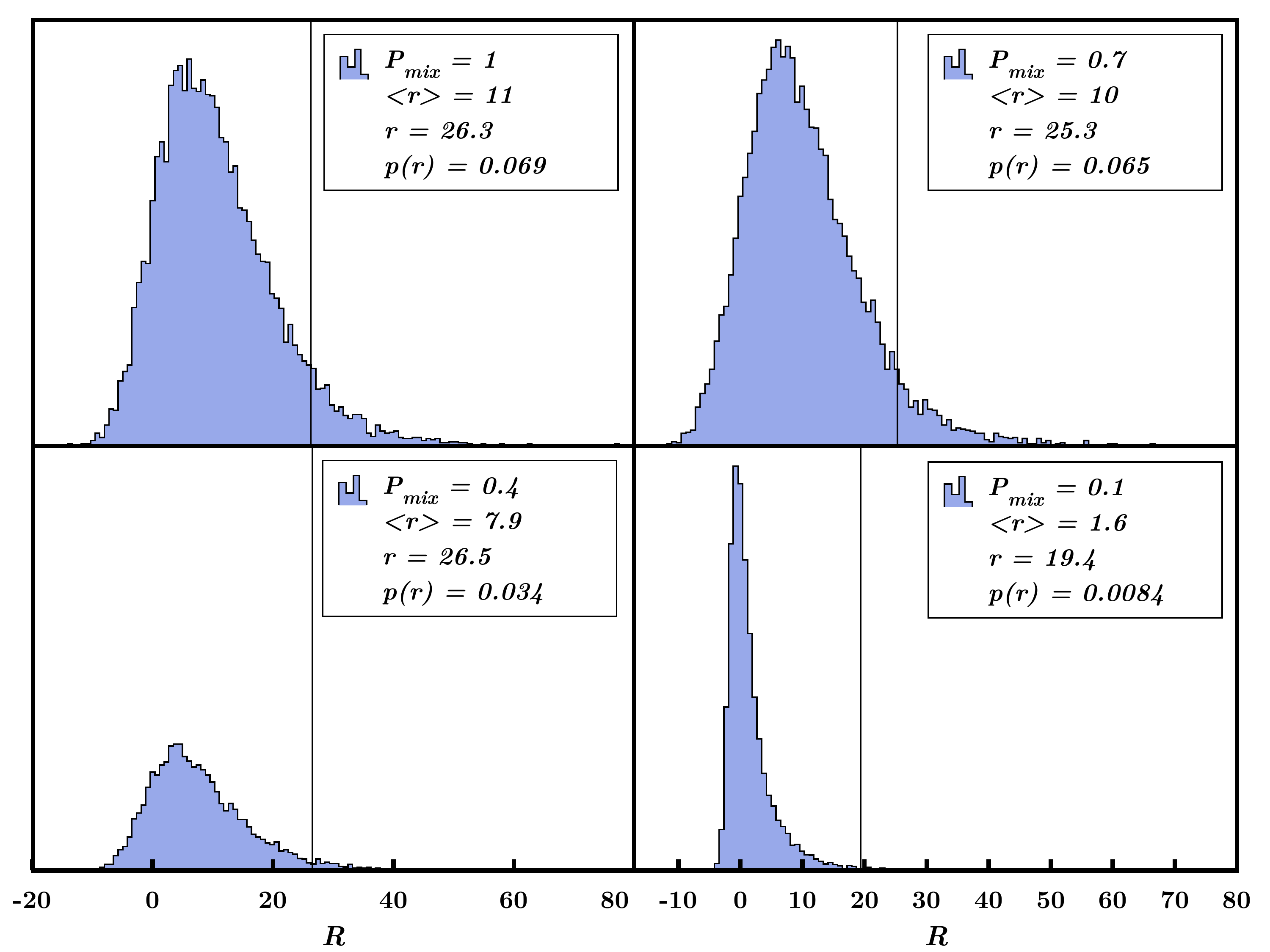}
    \caption{\rdist for the \Full catalog at $ E = \kev{2} $ when the scatter is distributed according to the \ALPmodel.}
    \label{fig:r_hist_full_x2_alp}
\end{figure}

A similar situation is found for the \Full catalog, where the measured signal is systematically higher than the expected one -- see \fref{fig:r_hist_full_x2_alp}.   The probability of such measurements reaches the $ 7\% $ level for $ \Pmix = 1 $.

The \BDSa, \BDSb and \Full catalogs all suggest that something different than \mixing is affecting the \Xray-to-optical luminosity ratio in a way that mimics the mixing effect.   As we already pointed out in \sref{sub:problem_xray_absorption}, \Xray absorption could be the cause.

\begin{figure}[htbp]
 \includegraphics[width=\imagewidth]{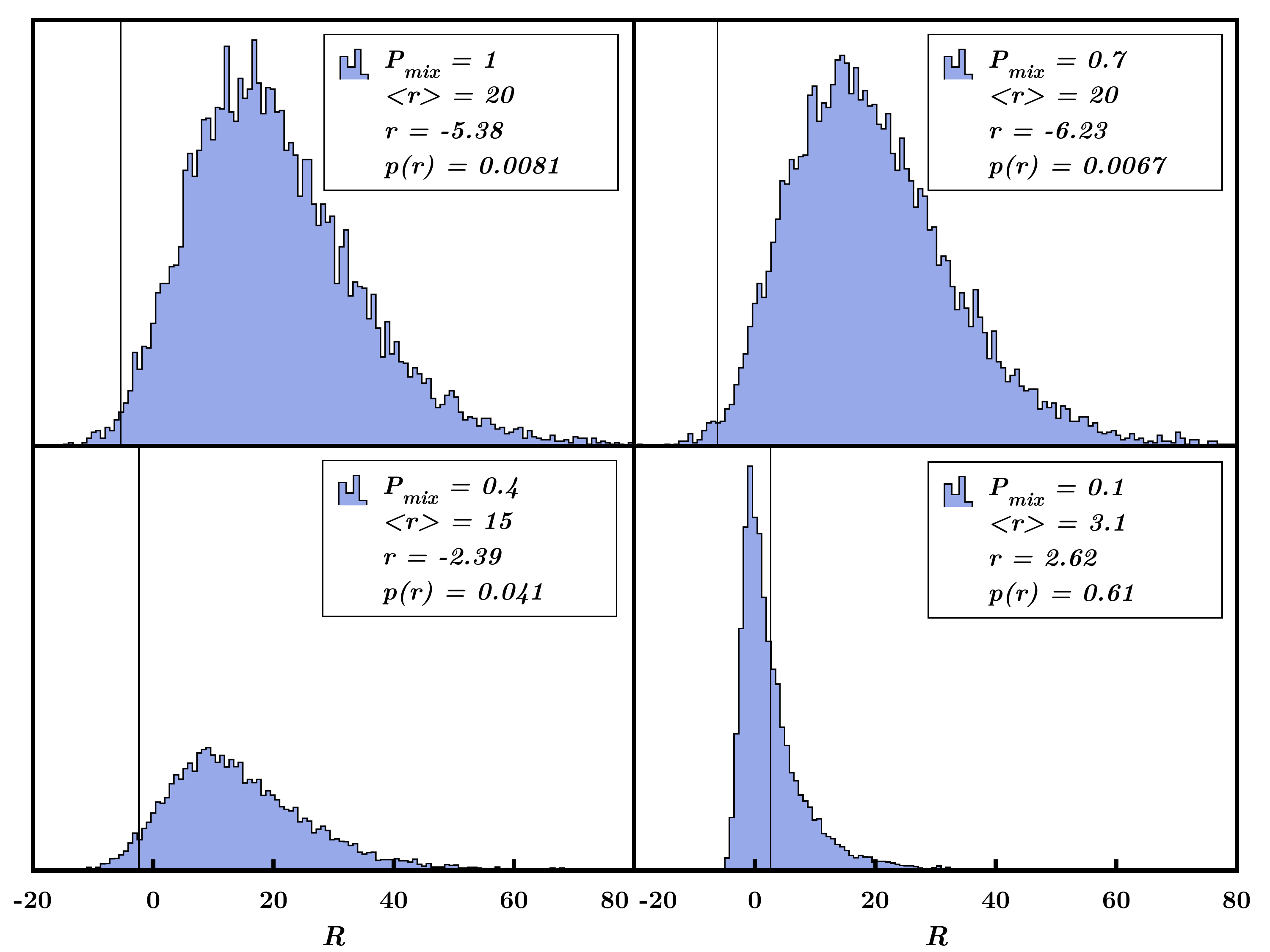}
    \caption{\rdist for the \HighG catalog at $ E = \kev{2} $ when the scatter is distributed according to the \ALPmodel.}
    \label{fig:r_hist_highg_x2_alp}
\end{figure}

The \HighG catalog, due to its low scatter variance, can potentially provide a statistically significant detection of \mixing.   This is evident from \fref{fig:r_hist_highg_x2_alp}, where the average outcome of the \rtest can be as high as $ \avg{r} = 20 $.   However, the measured signal never shows a significative preference for the \ALPmodel.   In fact, if favors the \Gaussmodel in most of the cases ($ \Pmix = 1, \, \pmixhigh, \, \pmixmiddle $).

\subsection{\Gaussmodel results}
\label{sub:gaussmodel_sim_results}
The measured $ r $-values are compatible with what expected from a Gaussian scatter only in the following cases:
\begin{enumerate}
   \item for the \HighG catalog, when $ E \geq \kev{2} $;
   \item for the \Full catalog, when $ E \geq \kev{4} $.
\end{enumerate}

\begin{figure}[htbp]
    \myfloatalign \includegraphics[width=\imagewidth]{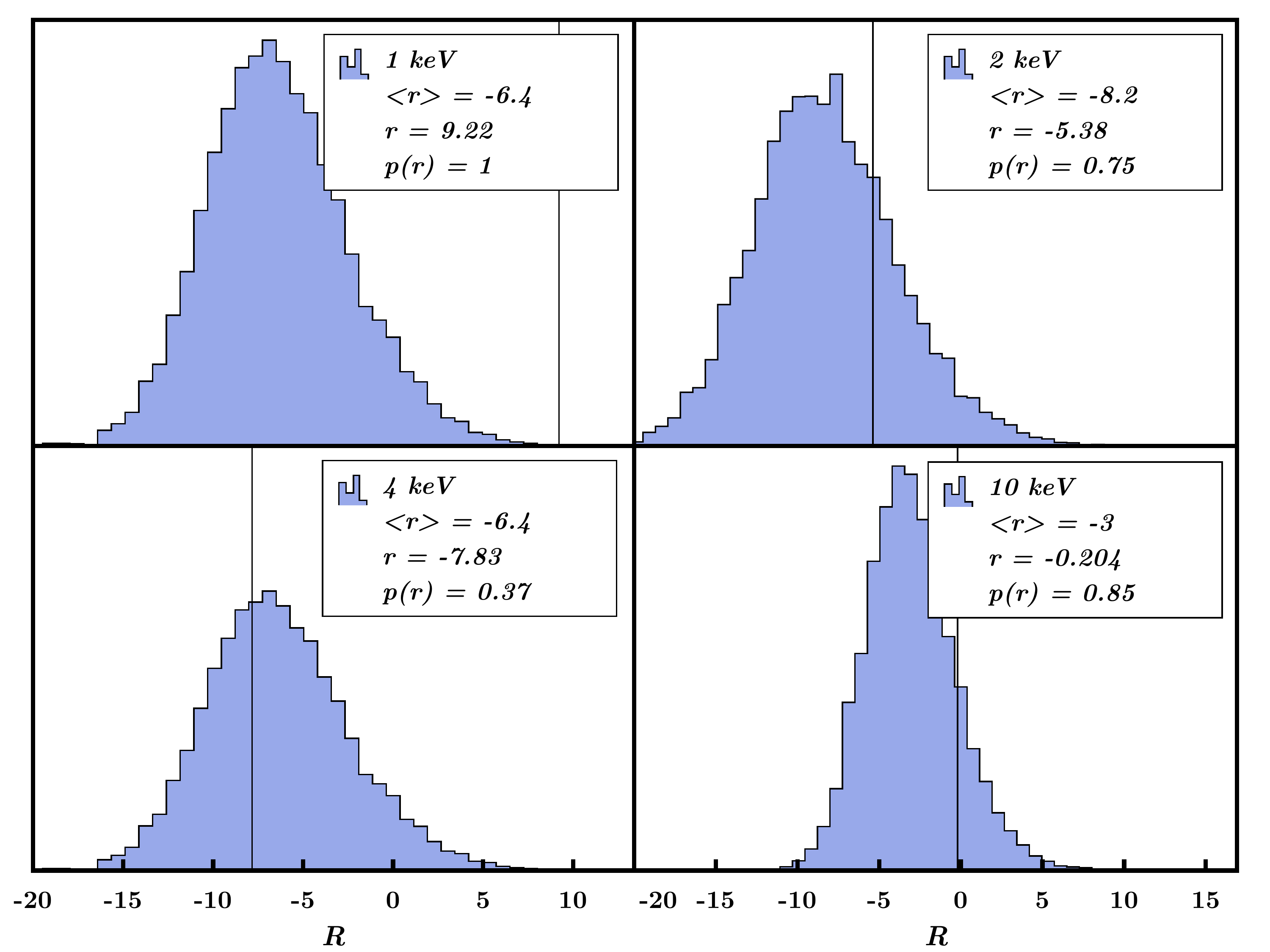}
    \caption{\rdist for the \HighG catalog at $ \Pmix = 1 $ for $ 4 $ different frequencies when the scatter is distributed according to the \Gaussmodel.}
    \label{fig:r_hist_highg_x4_gauss}
\end{figure}

This means that the \Gaussmodel performs well according to the \rtest where we expect \Xray absorption to be negligible.   An example of this is shown in \fref{fig:r_hist_highg_x4_gauss}, where we show the \rdist for the \HighG catalog for $ 4 $ different frequencies.   For frequencies higher than \kev{2}, the measured signal lies comfortably within the expected distribution of $ r $-values .

\subsection{\Shotmodel results}
\label{sub:shotmodel_sim_results}
\begin{figure}[htbp]
    \myfloatalign \includegraphics[width=\imagewidth]{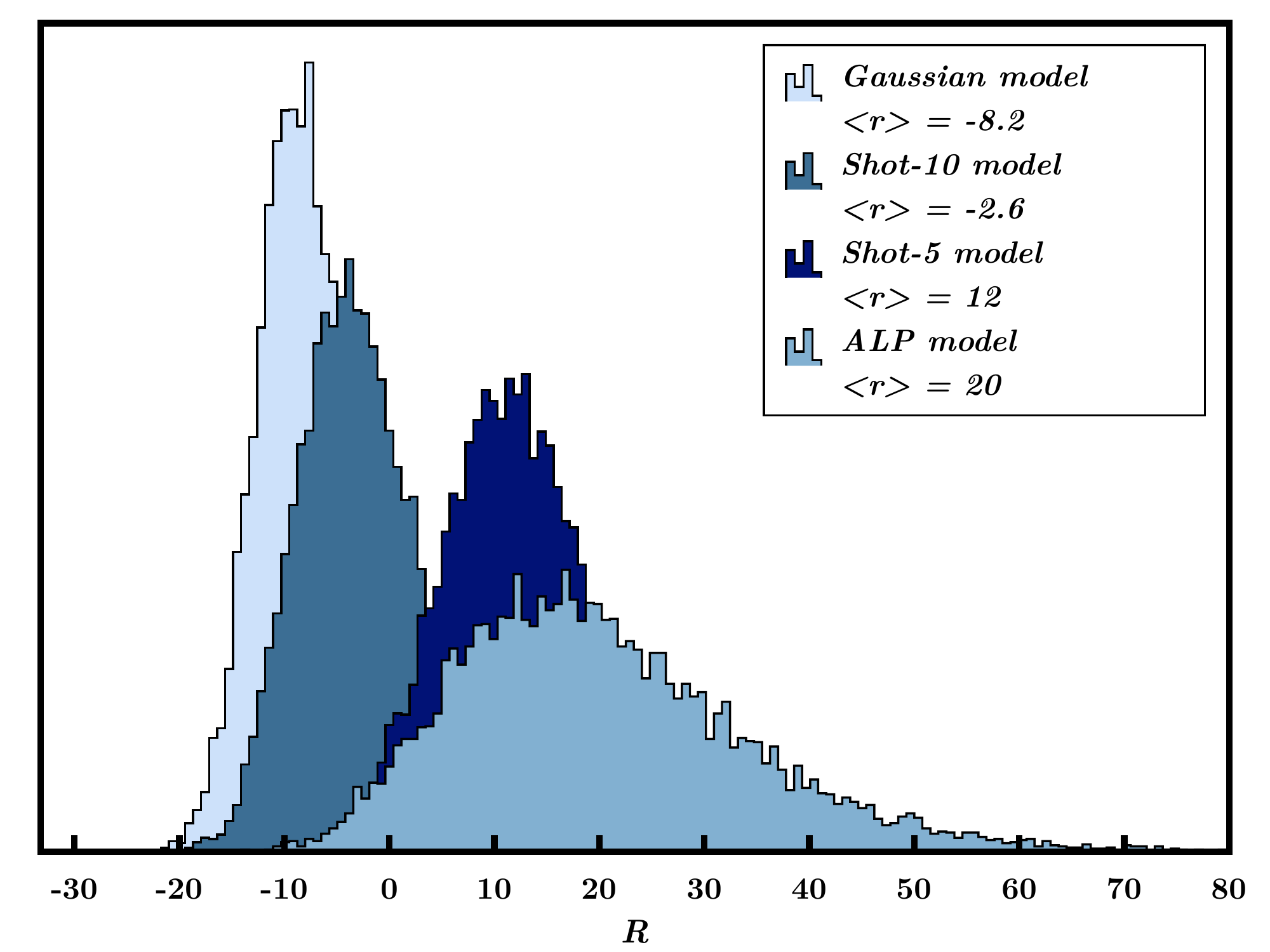}
    \caption{\rdist for the \HighG catalog at $ E = \kev{2} $ and $ \Pmix = 1 $ for $ 4 $ different models. The shot-noise models stay in between the Gauss and ALP-mixing models.}
    \label{fig:r_hist_highg_x2_shot5}
\end{figure}

The introduction of shot-noise on top of a Gaussian-distributed scatter can significantly alter the outcome of the \rtest if less than $ \avg{N}\!=\!20 $ \Xray photons are collected. The impact on the \rdist is a shift of its peak towards higher $ r $-values and a broadening of its width.   Moreover, when the number of collected photons is smaller than $ \sim\!10 $, shot-noise can even trick the \rtest into showing a preference for the ALP model (\ie a positive $ r $-value).   For example, for the \HighG catalog and when $ \avg{N} = 5 $, the \rtest yields an average value of $ 12 $, that is a $ \sim\!3.5 \sigma $ preference for the \ALPmodel.   All of this is clear from \fref{fig:r_hist_highg_x2_shot5}, where we show the \rdists for the \HighG catalog for 4 different models: \ALPmodel, \Gaussmodel, \Shotmodel with $ \avg{N}\!=\!5 $ and \Shotmodel with $ \avg{N}\!=\!10 $.

We do not expect the \Full and \HighG catalogs to be affected by shot-noise since all their AGNs are detected with far more than $ 20 $ \Xray counts.   However, a non-negligible number of AGNs in the \BDSa and \BDSb catalogs -- around the $ 10\% $ of the total, see \sref{sec:previous_analyses} -- have photon counts below $ 20 $.   Therefore, shot-noise may contribute to the high $ r $-values measured for those catalogs.   Nevertheless, the impact of shot noise is likely to be much smaller than that due to \Xray absorption.

\subsection{Goodness-of-fit tests}
\label{sec:Goodness-of-fit_tests}
We tested the measured scatter against the ALP-mixing, Gaussian and shot-noise models by making use of the \KS, Kuiper and Anderson-Darling tests.  The \KS and Kuiper tests tend to be more sensitive to the centers of the distribution, while the Anderson-Darling is weighted to emphasize the tails.   For every dataset, \Xray frequency and scatter model, we calibrated the goodness-of-fit tests by applying them to \nsimsgoftest simulated scatter samples.   \tref{tab:goodness_of_fits_tests} shows the statistical significances obtained by comparing the measured statistics to the simulated distributions.   Values lower than $ 0.05 $ imply a rejection of the model at the 2$\sigma$ level.
\begin{table*}[htbp]\footnotesize
 \begin{tabular}{l l |*{3}c|*{3}c|*{3}c|*{3}c|*{3}c|*{3}c }
    \multicolumn{2}{c}{} & \multicolumn{3}{|c|}{\kev{1}} & \multicolumn{3}{|c|}{\kev{1.5}} & \multicolumn{3}{|c|}{\kev{2}} & \multicolumn{3}{|c|}{\kev{4}} & \multicolumn{3}{|c|}{\kev{7}} & \multicolumn{3}{|c}{\kev{10}}\\
    \hline \hline
    \multirow{4}{*}{ALP}
    & \HighG  &0.17&0.15&0.0021&0.13&0.3&6.2e-4&0.061&0.22&4.2e-4&0.078&0.028&6.6e-4&0.061&0.18&0.0051&0.033&0.023&0.017\\
    & \Full   &0&0&0.03&2e-4&8e-5&0.064&0.028&0.0042&0.014&0.082&0.22&0.0026&0.024&0.067&0.008&0.036&0.046&0.026\\	 
    & \BDSa   &&&&&&&0.27&0.11&0.067&&&&&&&&&\\	 
    & \BDSb   &&&&&&&0.056&0.008&0.067&&&&&&&&&\\	 
    \hline
    \multirow{4}{*}{Gauss}
    & \HighG  &0.005&2.2e-4&0.015&0.035&0.004&0.017&0.005&7e-4&0.011&0.043&0.011&0.014&0.089&0.03&0.035&0.051&0.0039&0.071\\	 
    & \Full   &0&0&3.6e-4&0&0&0.001&4e-5&0&0.0031&0.06&0.0068&0.031&0.039&0.0053&0.046&0.073&0.0083&0.099\\	 
    & \BDSa   &&&&&&&0.062&0.029&1.6e-4&&&&&&&&&\\
    & \BDSb   &&&&&&&0.012&0.0035&3.2e-4&&&&&&&&&\\	 
    \hline
    \multirow{4}{*}{Shot}
    & \HighG  &0.064&0.039&0.0016&0.026&0.042&0.001&0.01&0.041&9e-4&0.019&0.11&7.8e-4&0.024&0.057&0.0047&0.014&0.0054&0.013\\	 
    & \Full   &0&0&0.0019&8e-5&0&0.0044&0.0084&7.6e-4&0.011&0.026&0.12&0.0026&0.0087&0.016&0.0059&0.018&0.014&0.021\\	 
    & \BDSa   &&&&&&&0.19&0.062&0.0019&&&&&&&&&\\
    & \BDSb   &&&&&&&0.032&0.0023&0.0021&&&&&&&&&\\	 
    \end{tabular}
    \caption{Statistical significance of the KS, Kuiper and AD statistics obtained for the various catalogs. For the \ALPmodel we assumed $ p_0 = 0.1 $ and $ \Pmix = 1 $; for the \Shotmodel we used $ \avg{N} = 5 $.}
\label{tab:goodness_of_fits_tests}
\end{table*}

The most obvious result of the goodness-of-fit tests is that none of the distributions does that well in reproducing the data.  Regardless of the analyzed model or dataset, most of the measured signals have a statistical significance smaller than $ 10\% $.  Every test fails at the 1$\sigma$ level, and only one data set is accepted in all three tests at the 2$\sigma$ level (the \BDSa set for the \ALPmodel).

The \rtest is most sensitive to the tails of the distributions, so we expect the AD test to be the best predictor of the $r$ value.   Indeed, this is the case: where the $r$ value is high, the AD test gives a very low value for the Gaussian distribution, and higher values for the \ALPmodel.  However, though the values are higher for the \ALPmodel, they remain far lower than would typically be expected.   This is consistent with what can be seen from the distributions in Fig. \ref{fig:cdf_bds} and \ref{fig:cdf_sdssxmm}. (Note that the figures show only the left tail.)

In the soft \Xray band, most of the tests yield very low significances, particularly in the \Full data set (the \Full catalog is never fit by any model for $ E \leq \kev{2} $.)  This suggests that none of the distributions well reflects the data and that something, such as \Xray absorption, is missing in the models.   This is generally improved when going to the harder \Xray bands or the better cleaned \HighG data set. 
This is not surprising, as the scatter in these cases tends to fall in between the \ALPmodel and Gaussian distributions.

These tests generally confirm what we have seen above:   preference for the \ALPmodel, as reflected in the $r$ statistic, should not be confused with the data being consistent with this model.   Indeed, there are strong indications that the simple models we consider fail to explain the data, particularly in the soft \Xray band.


\section{Conclusions}
\label{sec:conclusions}

The ratios of luminosities of high redshift objects, introduced by \citet{burrage:2009a}, offers a new avenue to explore the possibility of strong mixing between photons and axions, which in principle can have significant power to constrain these models.  However, much effort must be taken to ensure a homogenous sample of objects, so that any scatter is due to the coupling to axions.    

Here, we have reproduced the analysis of \citeauthor{burrage:2009a} studying scatter of the empirical relation between the optical/UV and \Xray monochromatic luminosities of AGNs. In addition to the AGN catalogs already analysed by them -- \BDSa and \BDSb~--, we considered two samples from the \sdssxmm Quasar Survey \citep{young:2009b}: \Full-\nagnFull and \HighG-\nagnHigh.  These data sets have multi-wavelength information which has allowed a more thorough investigation of the model constraints. 

We ran statistical analyses on $ 18 $ combinations of optical/UV and \Xray monochromatic luminosities and found no compelling evidence for the presence of ALPs. Whenever we found a signal compatible with the presence of ALPs, it was coming either from the soft \Xray band, where absorption is more likely to happen, or from the \Full catalog, which is thought to include more absorbed AGNs. Moreover, the \ALPmodel fails to reproduce the features of the scatter even where it is preferred over its competing model, the simple \Gaussmodel.

By means of simulations, we calculated the distribution of the \rstat used by \citeauthor{burrage:2009a} to discriminate between the \ALPmodel and the \Gaussmodel. We found out that actual data yields values of the \rstat which are either significantly higher (\BDSb and \Full catalogs) or lower (\HighG catalog) than what we would typiclally expect if \mixing were taking place.
Stated differently, we see ALP evidence where there should not be sensitivity, and we do not see evidence in the majority of data sets which should be most sensitive.  

This behaviour suggests that another source of scatter, such as \Xray absorption, is taking place and casts doubts on the suitability of the \rstat alone to estimate \mixing.  This is supported by a detailed examination of the low luminosity sources which dominate the statistical tests.  We found that the \BDSa and \BDSb catalogs respectively contain one and two AGNs which are known to be \Xray-weak independently of \mixing. Upon removing these sources, the value of the \rstat significantly drops in both cases, as does the preference for the \ALPmodel over the \Gaussmodel.

Thus, while this new method for observing photon-ALP mixing is in principle very powerful, considerable care must be taken.  Given the many ways that scatter can be introduced into the relations between the low and high energy luminiosities of objects, evidence for  mixing must be closely examined that it fits the expected signal.   In particular, the scatter should follow the expected PDF and be independent of the energy of the high energy photons.   Ideally, it should also be observed in multiple classes of objects, where their intrinsic scatter have different physical origins.

This also highlights the importance of finding classes of objects where the intrinsic scatter is small or well understood.   Since the scatter from mixing is well understood and fixed, if any sample is observed with less scatter than expected from mixing, then mixing can be ruled out, or at the very least, the probability of mixing, \Pmix, can be constrained.

\section*{Acknowledgements}
\label{sec:ack}
We thank Monica Young for providing us with the monochromatic luminosities of the Quasars in the SDSS/Xmm-Newton Quasar Survey and Douglas Shaw for many useful remarks.
In addition, we acknowledge many useful conversations with Clare Burrage, Heather Campbell, Timothy Clemson, Alejo Martinez-Sansigre, Bob Nichol, Kathy Romer and Daniel Thomas.

\bibliography{my_bibliography}

\end{document}